\documentclass[11pt,preprint]{aastex}
\usepackage{emulateapj5,apjfonts}

\renewcommand{\tabcolsep}{3pt}

\newcommand{\NH}{{$N_{\rm H}$}}
\newcommand{\LX}{{$L_{\rm X}$}}

\newcommand{\LHa}{$L_{{\rm H} \alpha}$}

\newcommand{\eps}{ergs s$^{-1}$}
\newcommand{\pcm}{cm$^{-2}$}

\newcommand{\chandra}{{\it Chandra}}
\newcommand{\asca}{{\it ASCA}}

\newcommand{\rosat}{{\it ROSAT}}

\newcommand{\HST}{{\it HST}}

\newcommand{\gtsima}{$\; \buildrel > \over \sim \;$}
\newcommand{\simgt}{\lower.5ex\hbox{\gtsima}}
\newcommand{\ltsima}{$\; \buildrel < \over \sim \;$}
\newcommand{\simlt}{\lower.5ex\hbox{\ltsima}}

\slugcomment{Submitted to {\it The Astrophysical Journal}}

\lefthead{Terashima et al.}

\righthead{}

\begin{document}

\title{Chandra Snapshot Observations of Low-Luminosity AGNs with a Compact Radio Source}

\author{Yuichi Terashima\altaffilmark{1} and Andrew S. Wilson\altaffilmark{2}}

\affil{Astronomy Department, University of Maryland, College Park, MD 20742}

\altaffiltext{1}{Institute of Space and Astronautical Science, 3-1-1 Yoshinodai, Sagamihara, Kanagawa 229-8510, Japan}

\altaffiltext{2}{Adjunct Astronomer, Space Telescope Science Institute, 3700
San Martin Drive, Baltimore, MD 21218}

\begin{abstract}

The results of {\chandra} snapshot observations of 11 LINERs
(Low-Ionization Nuclear Emission-line Regions), three low-luminosity
Seyfert galaxies, and one \ion{H}{2}-LINER transition object are
presented. Our sample consists of all the objects with a flat or
inverted spectrum compact radio core in the VLA survey of 48
low-luminosity AGNs (LLAGNs) by Nagar et al. (2000). An X-ray nucleus
is detected in all galaxies except one and their X-ray luminosities
are in the range $5\times10^{38}$ to $8\times10^{41}$ {\eps}.  The
X-ray spectra are generally steeper than expected from thermal
bremsstrahlung emission from an advection-dominated accretion flow
(ADAF). The X-ray to H$\alpha$ luminosity ratios for 11 out of 14
objects are in good agreement with the value characteristic of LLAGNs
and more luminous AGNs, and indicate that their optical emission lines
are predominantly powered by a LLAGN. For three objects, this ratio is
less than expected.  Comparing with properties in other wavelengths,
we find that these three galaxies are most likely to be heavily
obscured AGN.  We use the ratio $R_{\rm X} = \nu L_\nu$(5 GHz)/$L_{\rm
X}$, where {\LX} is the luminosity in the 2--10 keV band, as a measure
of radio loudness. In contrast to the usual definition of radio
loudness ($R_{\rm O} = L_{\nu}$(5 GHz)/$L_{\nu}$(B)), $R_{\rm X}$ can
be used for heavily obscured ({\NH} \simgt $10^{23}$ {\pcm}, $A_{\rm
V}>50$ mag) nuclei. Further, with the high spatial resolution of
{\chandra}, the nuclear X-ray emission of LLAGNs is often easier to
measure than the nuclear optical emission. We investigate the values
of $R_{\rm X}$ for LLAGNs, luminous Seyfert galaxies, quasars and
radio galaxies and confirm the suggestion that a large fraction of
LLAGNs are radio loud.

\end{abstract}


\section{Introduction}

Low-Ionization nuclear emission-line regions (LINERs; Heckman 1980)
are found in many nearby bright galaxies (e.g., Ho, Filippenko, \&
Sargent 1997a). Extensive studies at various wavelengths have shown
that type 1 LINERs (LINER 1s, i.e., those galaxies having broad
H$\alpha$ and possibly other broad Balmer lines in their nuclear
optical spectra) are powered by a low-luminosity AGN (LLAGN) with a
bolometric luminosity less than $\sim10^{42}$ {\eps} (Ho et al. 2001;
Terashima, Ho, \& Ptak 2000a; Ho et al. 1997b). On the other hand, the
energy source of LINER 2s is likely to be heterogeneous.  Some LINER
2s show clear signatures of the presence of an AGN, while others are
most probably powered by stellar processes, and the luminosity
ratio {\LX}/{\LHa} can be used to discriminate between these different
power sources (e.g., P\'erez-Olea \& Colina 1996;
Maoz et al. 1998; Terashima et al. 2000b).  It is interesting to note
that currently there are only a few LINER 2s known to host an obscured
AGN (e.g., Turner et al. 2001). This paucity of obscured AGN in LINERs
may indicate that LINER 2s are not simply a low-luminosity extension
of luminous Seyfert 2s, which generally show heavy obscuration with a
column density averaging {\NH} $\sim$ $10^{23}$ {\pcm} (e.g., Turner
et al. 1997). Alternatively, biases against finding heavily obscured
LLAGNs may be important. For example, objects selected through optical
emission lines or X-ray fluxes are probably biased in favor of less
absorbed ones, even if one uses the X-ray band above 2 keV.

  In contrast, radio observations, particularly at high frequency, are
much less affected by absorption. Although an optical spectroscopic
survey must first be done to find the emission lines characteristic of
a LLAGN, follow up radio observations can clarify the nature of the
activity. For example, VLBI observations of some LLAGNs have revealed
a compact nuclear radio source with $T_{\rm b}>10^8$ K, which is an
unambiguous indicator of the presence of an active nucleus and cannot
be produced by starburst activity (e.g., Falcke et al. 2000; Ulvestad
\& Ho 2001). A number of surveys of Seyfert galaxies at sub arcsecond
resolution have been made with the VLA (Ulvestad \& Wilson 1989 and
references therein; Kukula et al. 1995; Nagar et al. 1999; Thean et
al. 2000; Schmitt et al.  2001; Ho \& Ulvestad 2001) and other
interferometers (Roy et al. 1994; Morganti et al. 1999), but much less
work has been done on the nuclear radio emission of LINERs. Nagar et
al. (2002) have reported a VLA 2 cm radio survey of all 96 LLAGNs
within a distance of 19 Mpc. These LLAGNs come from the Palomar
spectroscopic survey of bright galaxies (Ho et al. 1997a). As a pilot
study of the X-ray properties of LLAGNs, we report here a {\chandra}
survey of a subset, comprising 15 galaxies, of Nagar et al's (2002)
sample. Fourteen of these galaxies have a compact nuclear radio core
with a flat or inverted radio spectrum (Nagar et al. 2000). We have
detected 13 of the galactic nuclei with {\chandra}. We also examine
the ``radio loudness'' of our sample and compare it with other classes
of AGN. A new measure of ``radio loudness'' is developed, in which the
5 GHz radio luminosity is compared with the 2--10 keV X-ray luminosity
($R_{\rm X}$=$\nu L_{\nu}$ (5 GHz)/$L$(2--10 keV)) rather than with
the B-band optical luminosity ($R_{\rm O}= L_{\nu}$(5
GHz)/$L_{\nu}$(B)), as is usually done. $R_{\rm X}$ has the advantage
that it can be measured for highly absorbed nuclei ({\NH} up to
several times $10^{23}$ {\pcm}) which would be totally obscured
($A_{\rm V}$ up to a few hundred mag for the Galactic gas to dust ratio)
at optical wavelengths, and that the compact, hard X-ray source in a
LLAGN is less likely to be confused with emission from stellar-powered
processes than is an optical nucleus.

  This paper is organized as follows. The sample, observations, and
data reduction are described in section 2. Imaging results and X-ray
source detections are given in section 3. Section 4 presents spectral
results. The power source, obscuration in LLAGNs, and radio loudness
of LLAGNs are discussed in section 5. Section 6 summarizes the
findings. We use a Hubble constant of $H_0=75$ km s$^{-1}$ Mpc$^{-1}$
and a deceleration parameter of $q_0=0.5$ throughout this paper.

\section{The Sample, Observations, and Data Reduction}

Our sample is based on the VLA observations by Nagar et
al. (2000). Their sample of 48 objects consists of 22 LINERs, 18
transition objects, which show optical spectra intermediate between
LINERs and \ion{H}{2} nuclei, and eight low-luminosity Seyferts
selected from the optical spectroscopic survey of Ho et
al. (1997a). The sample is the first half of a distance-limited sample
of LLAGNs (Nagar et al. 2002), as described in section 1.

  We selected 14 objects showing a flat to inverted spectrum radio
core ($\alpha \ge -0.3$, $S_{\nu}\propto \nu^{\alpha}$) according to
Nagar et al.'s (2000) comparison with longer wavelength radio data
published in the literature. One object (the LINER 2 NGC 4550) has a
flat spectrum radio source at a position significantly offset from the
optical nucleus. This object was added as an example of a LINER
without a detected radio core. The target list and some basic data for
the final sample are summarized in Table 1. The distances are taken
from Tully (1988) in which $H_0=75$ km s$^{-1}$ Mpc$^{-1}$ is assumed.
The sample consists of seven LINER 1s, four LINER 2s, two Seyfert 1s,
one Seyfert 2, and one transition 2 object. 12 out of these 15 objects
have been observed with the VLBA and high brightness temperature
($T_{\rm b}>10^7$ K) radio cores were detected in all of them (Falcke
et al. 2000; Ulvestad \& Ho 2001; Nagar et al. 2002). Therefore, these
objects are strong candidates for AGNs.  Results of archival /
scheduled {\chandra} observations of more LINERs with a compact
flat/inverted spectrum radio core found by Nagar et al. (2002) will be
presented in a future paper.

  A log of {\chandra} observations is shown in Table 1. The exposure
time was typically two ksec each. All the objects were observed at or
near the aim point of the ACIS-S3 back-illuminated CCD chip.  Eight
objects were observed in our Guaranteed Time Observation program and
the rest of the objects were taken from the {\chandra} archives. The
eight objects were observed in 1/8 sub-frame mode (frame time 0.4 s)
to minimize effects of pileup (e.g., Davis 2001). 1/2 sub-frame modes
were used for three objects.  CIAO 2.2.1 and CALDB 2.7 were used to
reduce the data. In the following analysis, only events with {\asca}
grades 0, 2, 3, 4, 6 (``good grades'') were used. For spectral
fitting, XSPEC version 11.2.0 was employed.

\section{X-ray Images and Source Detection}

\subsection{Source Detection}

An X-ray nucleus is seen in all the galaxies except for NGC 4550 and
NGC 5866. Some off-nuclear sources are also seen in some fields.  The
source detection algorithm ``wavdetect'' in the CIAO package was
applied to detect these nuclear and off-nuclear sources, where a
detection threshold of $10^{-6}$ and wavelet scales of 1, $\sqrt 2$,
2, $2\sqrt 2$, 4, $4\sqrt 2$, 8, $8\sqrt 2$, and 16 pixels were
used. Source detections were performed in the three energy bands
0.5--8 keV (full band), 0.5--2 keV (soft band), and 2--8 keV (hard
band). The resulting source lists and raw images were examined by eye
to exclude spurious detections.  The source and background counts were
also determined by manual photometry and compared with the results of
wavdetect. In the few cases that the two methods gave discrepant
results, we decided to use the results of the manual photometry after
inspection of the raw images. Some sources were detected in only one
or two energy bands. In such cases, we calculated the upper limits on
the source counts in the undetected band(s) at the 95\% confidence
level by interpolating the values in Table 2 of Kraft, Burrows, \&
Nousek (1991).

Table 2 shows the positions, detected counts, band ratios (hard/soft
counts), fluxes and luminosities in the 2--10 keV band of the nuclear
sources. The same parameters for off-nuclear sources with
signal-to-noise ratios greater than three are summarized in Table 3.
For bright objects ($>$ 40 counts), the fluxes were measured by
spectral fits presented in the next section. For faint objects,
fluxes were determined by assuming the Galactic absorption column
density and power law spectra, with photon indices determined from the
band ratios. When only lower or upper limits on the band ratio were
available, a photon index of 2 was assumed if the limit is consistent
with $\Gamma=2$. When the band ratio is inconsistent with $\Gamma=2$,
the upper or lower limit is used to determine the photon index.
Luminosities were calculated only if the source is spatially inside
the optical host galaxy as indicated by comparing the position with
the optical image of the Digitized Sky Survey. Possible
identifications for off-nuclear sources are also given in the last
column of Table 3.

The positions of the X-ray nuclei coincide with the radio core
positions to within the positional accuracy of {\chandra}. The nominal
separations between the X-ray and radio nuclei are in the range 0.05 --
0.95$^{\prime\prime}$.

\subsection{X-ray Morphology of the Nuclear Region}

Inspection of the images shows that the nucleus in most objects
appears to be unresolved, while some objects show faint extended
emission. The soft and hard band images of the nuclear regions of NGC
3169 and NGC 4278 are shown in Fig 1 as examples of extended emission.
In the soft band image of NGC 3169, emission in the nuclear region
extending $\sim10$ arcsec in diameter is clearly visible, while the
nucleus itself is not detected (Table 2).  About 25 counts were
detected within a 10 pixel (4.9 arcsec) radius in the $0.5-2$ keV
band. This extended emission is not seen in the hard band. The soft
band image of NGC 4278 consists of a bright nucleus and a faint
elongated feature with a length of $\sim50^{\prime\prime}$ and a
position angle of $\sim70^{\circ}$.  The hard band image is 
unresolved.

  The nuclear regions (10 arcsec scale) of the other galaxies look
compact to within the current photon statistics.  More extended
diffuse emission at larger scales ($>0.5^{\prime}$) is seen in a few
objects.  The nuclei of NGC 2787 and NGC 4203 are embedded in soft
diffuse emission with diameters of $\sim30^{\prime \prime}$ and
$50^{\prime \prime}$, respectively.  NGC 4579 shows soft diffuse
emission with a similar morphology to the circumnuclear star forming
ring, in addition to a very bright nucleus (see also Eracleous et
al. 2002).  NGC 4565 shows extended emission along the galactic
plane. NGC 5866 has soft extended emission $\approx 0.5^{\prime}$ (2
kpc) in diameter and no X-ray nucleus is detected (Table 2). Any
diffuse emission associated with the other galaxies is much fainter.

\begin{figure*}[t]
\noindent
\includegraphics[scale=0.49,angle=-90]{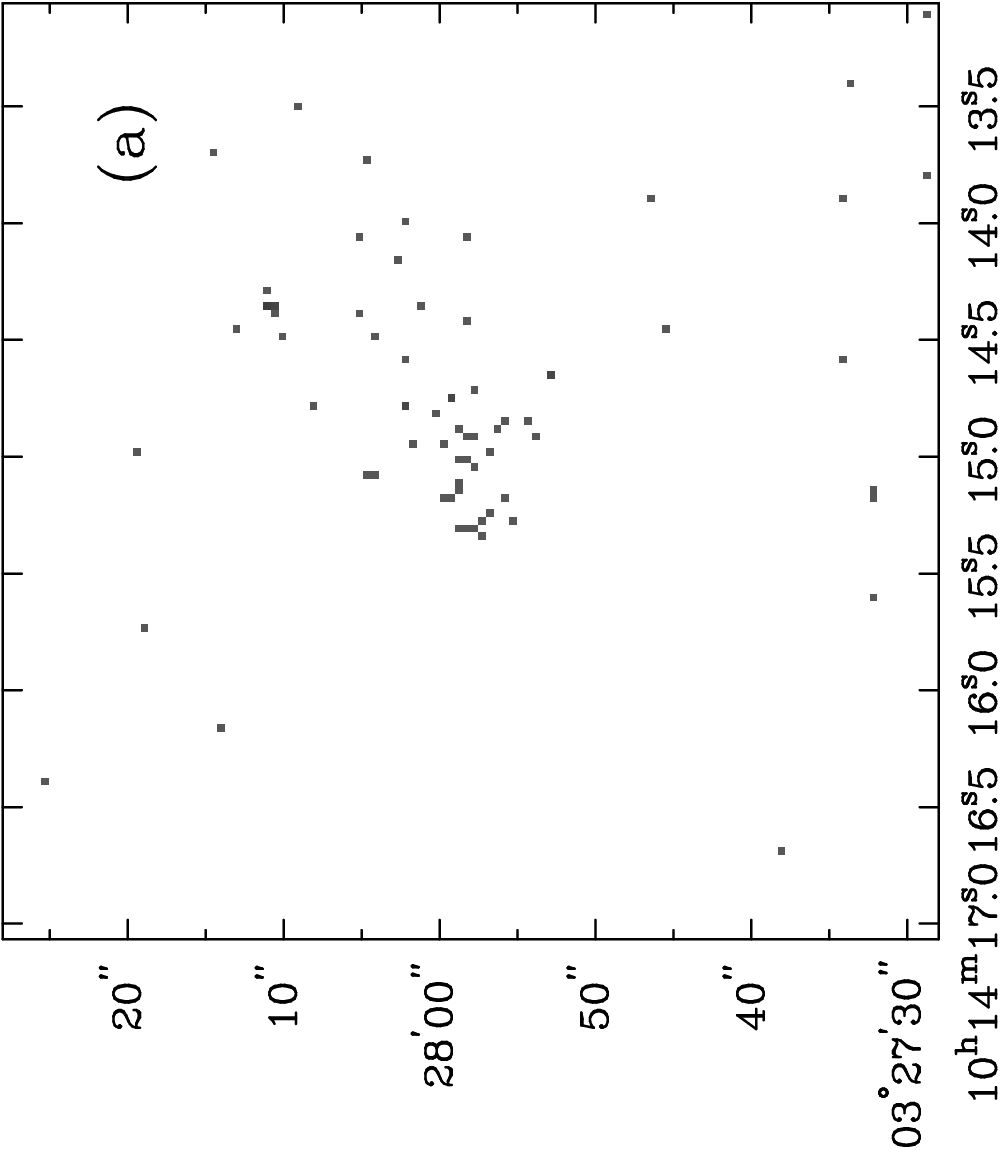}
\includegraphics[scale=0.49,angle=-90]{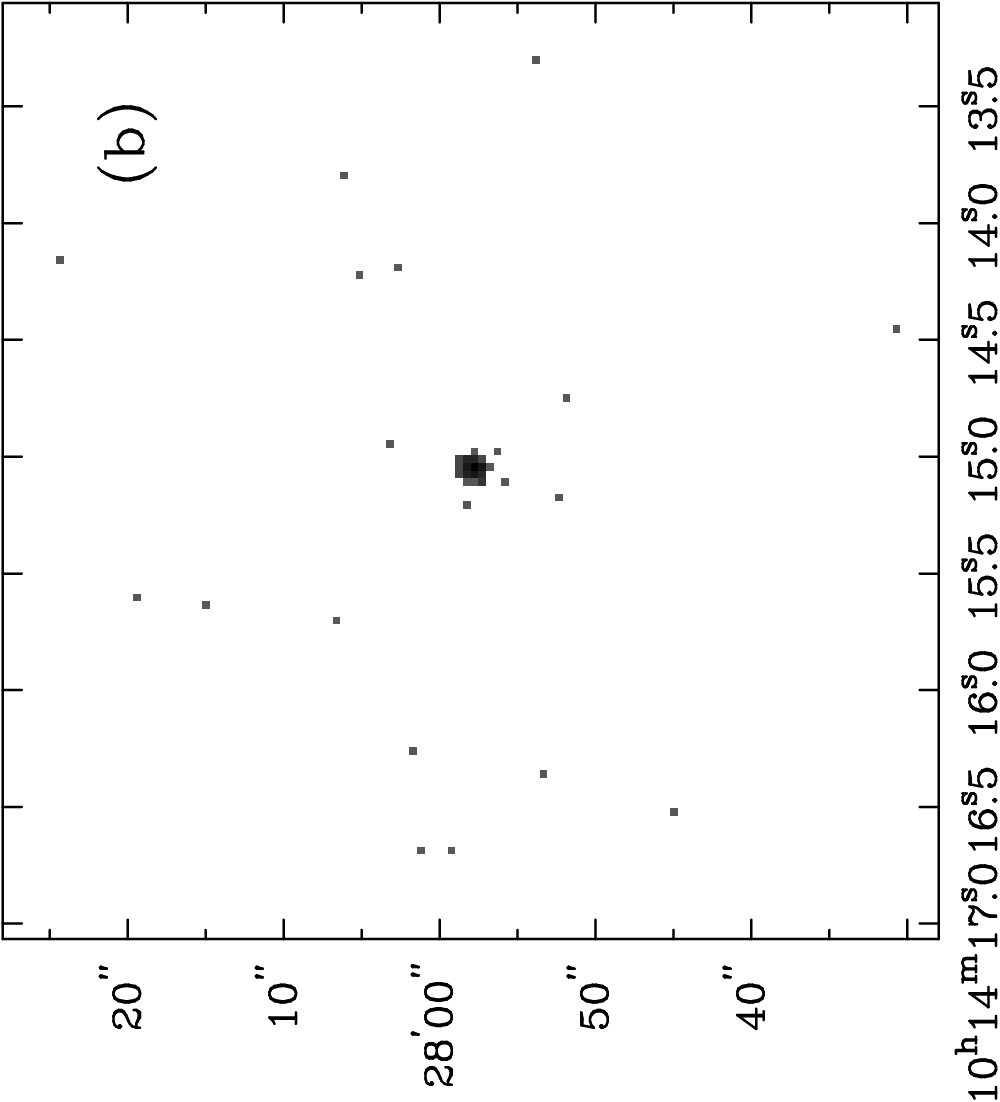}

\vspace{5mm}

\noindent
\includegraphics[scale=0.49,angle=-90]{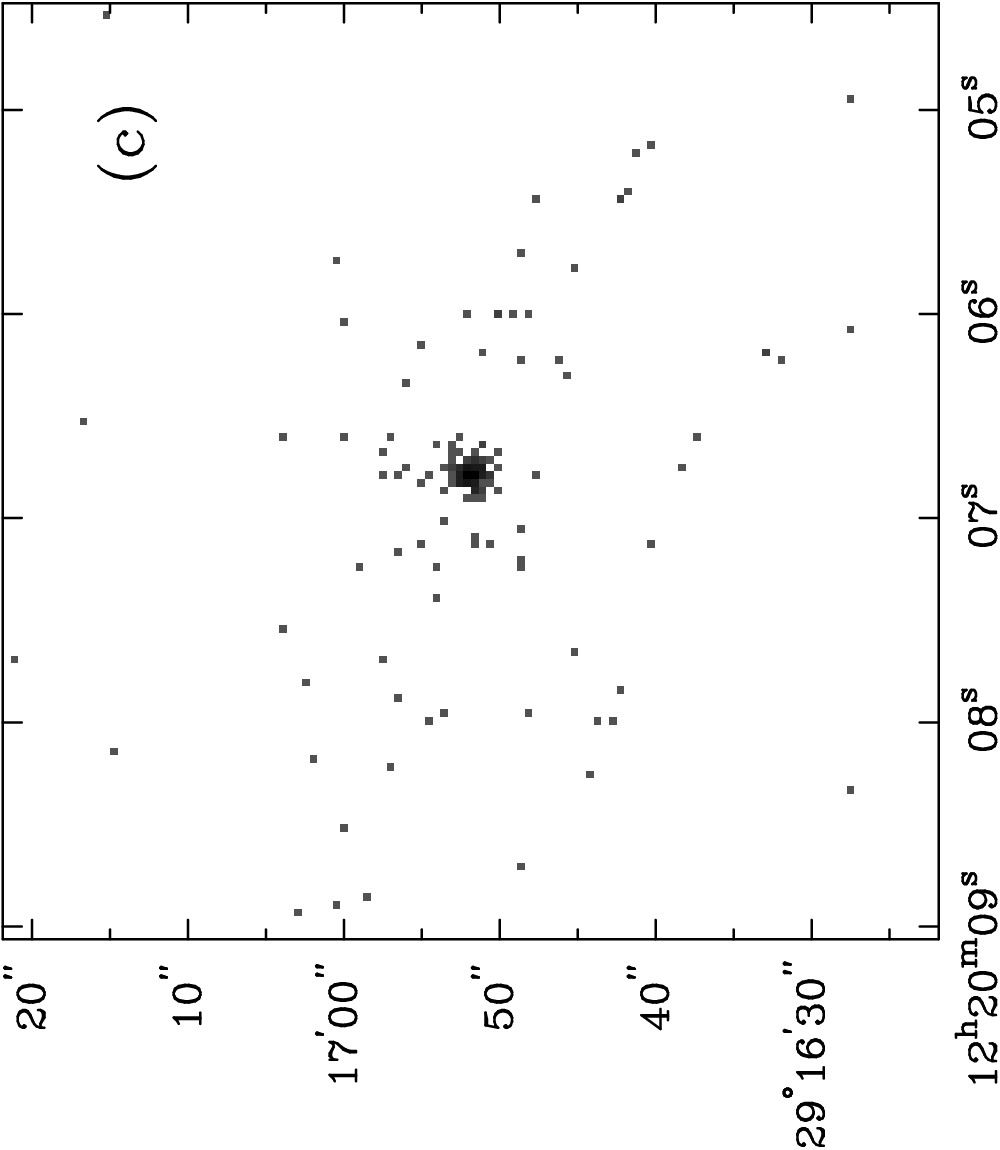}
\hspace{0.5mm} \includegraphics[scale=0.49,angle=-90]{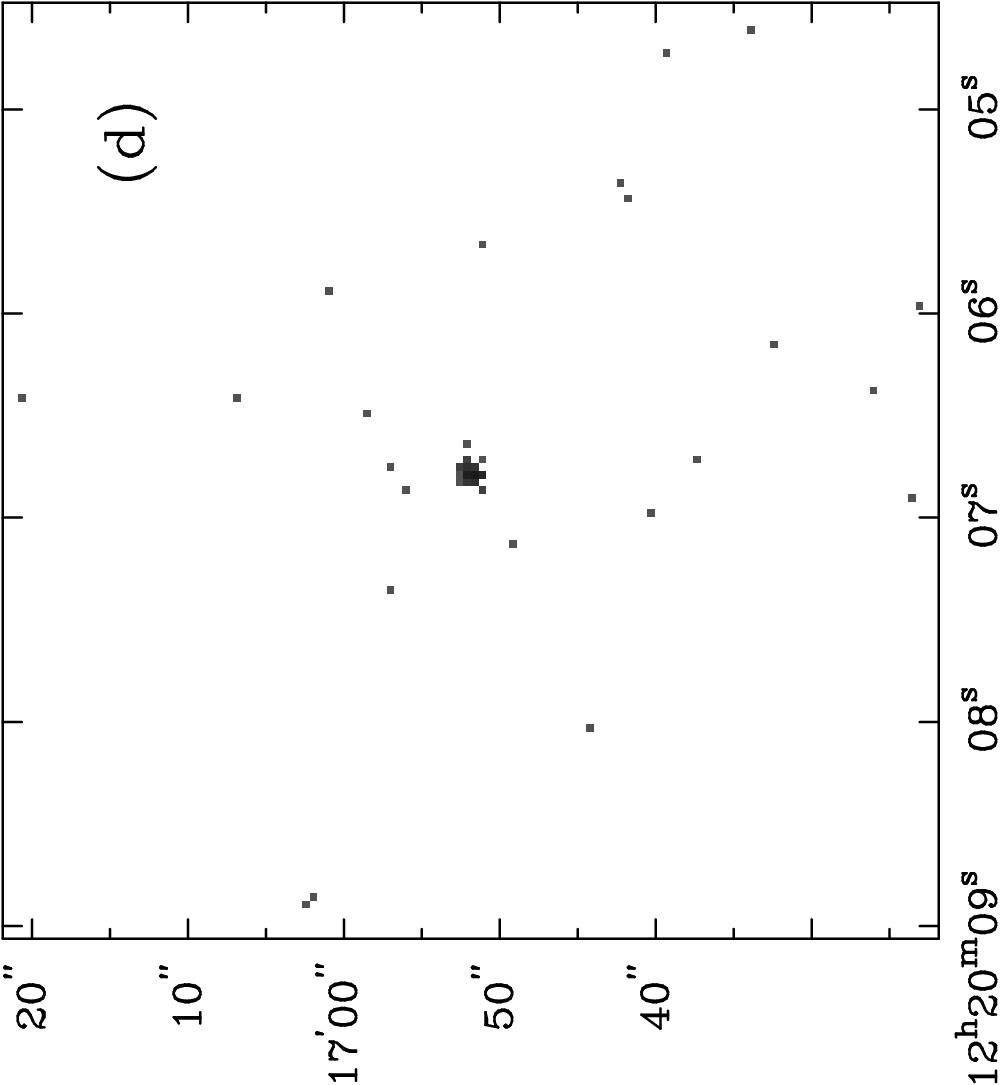}
\figcaption[]{
Examples of {\chandra} images.
({\it a}) NGC 3169 (0.5--2 keV). 
({\it b}) NGC 3169 (2--8 keV). 
({\it c}) NGC 4278 (0.5--2 keV). 
({\it d}) NGC 4278 (2--8 keV). 
\label{fig-1}
}
\end{figure*}

\section{X-ray Spectra}

Spectral fits were performed for the relatively bright objects ---
those with $>$ 40 detected counts in the 0.5--8 keV band.  The
spectrum of one fainter object (NGC 4548) showing a large (=hard)
hardness ratio was also fitted.  Some objects are so bright that
pileup effects are significant. Column 7 of Table 1 gives the count
rates per CCD read-out frame time and can be used to estimate the
significance of pileup.  In the two objects NGC 3147 and NGC 4278, the
pileup is mild and we corrected for the effect by applying the pileup
model implemented in XSPEC, where the grade morphing parameter
$\alpha$ was fixed at 0.5 (after initially treating it as a free
parameter [Davis 2001] since $\alpha$ is not well constrained). The
pileup effect for the three objects with the largest count rate per
frame (NGC 4203, NGC 4579, and NGC 5033) is serious and we did not
attempt detailed spectral fits. Instead, we use the spectra and fluxes
measured with {\asca} for these three objects (Terashima et al. 2002b
and references therein) in the following discussions. We confirmed
that the nuclear X-ray source dominates the hard X-ray emission within
the beam size of {\asca} (see Appendix).  The other objects in the
sample are faint enough to ignore the effects of pileup.

X-ray spectra were extracted from a circular region with a radius
between 4 pixels (2.0$^{\prime\prime}$; for faint sources) and 10
pixels (4.9$^{\prime\prime}$; for bright sources) depending on source
brightness. Background was estimated using an annular region centered
on the target. A maximum-likelihood method using the C-statistic (Cash
1979) was employed in the spectral fits. In the fit with the
C-statistic, background cannot be subtracted, so we added a background
model (measured from the background region) 

\vspace{1cm}

\begin{figure*}[t]
\noindent
\includegraphics[scale=0.49,angle=-90]{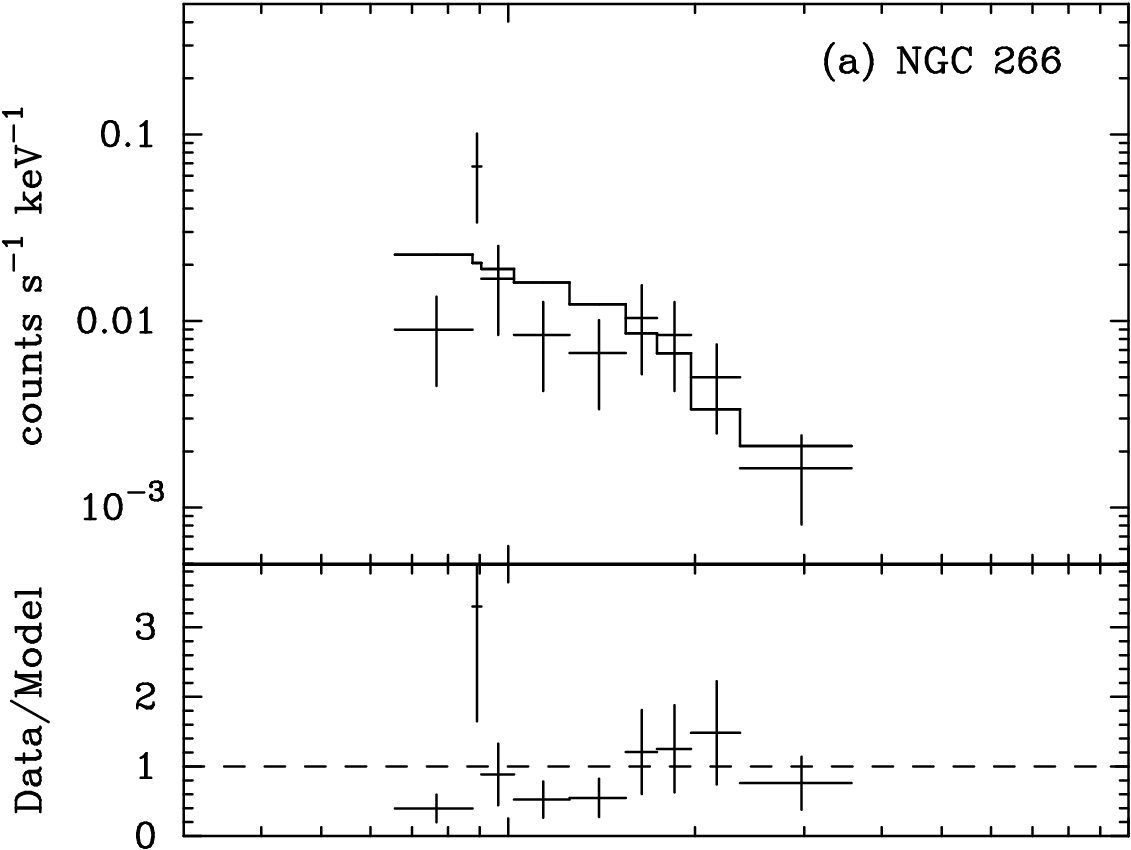}
\includegraphics[scale=0.49,angle=-90]{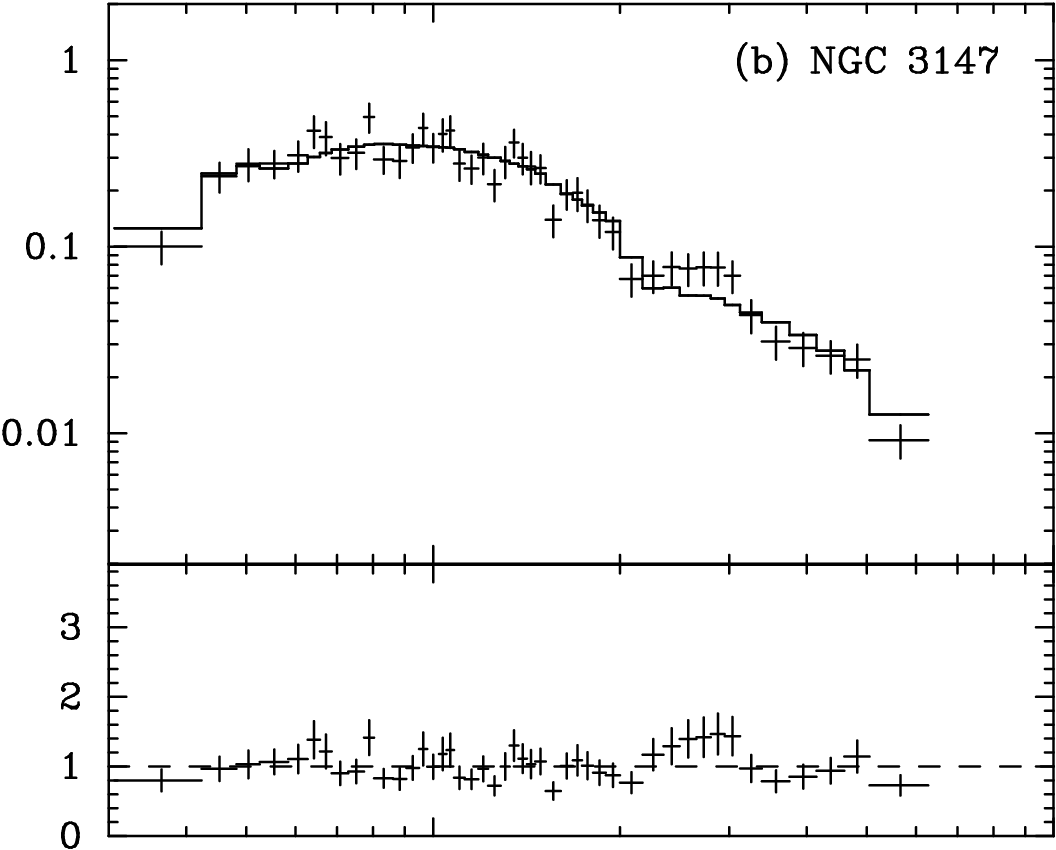}
\includegraphics[scale=0.49,angle=-90]{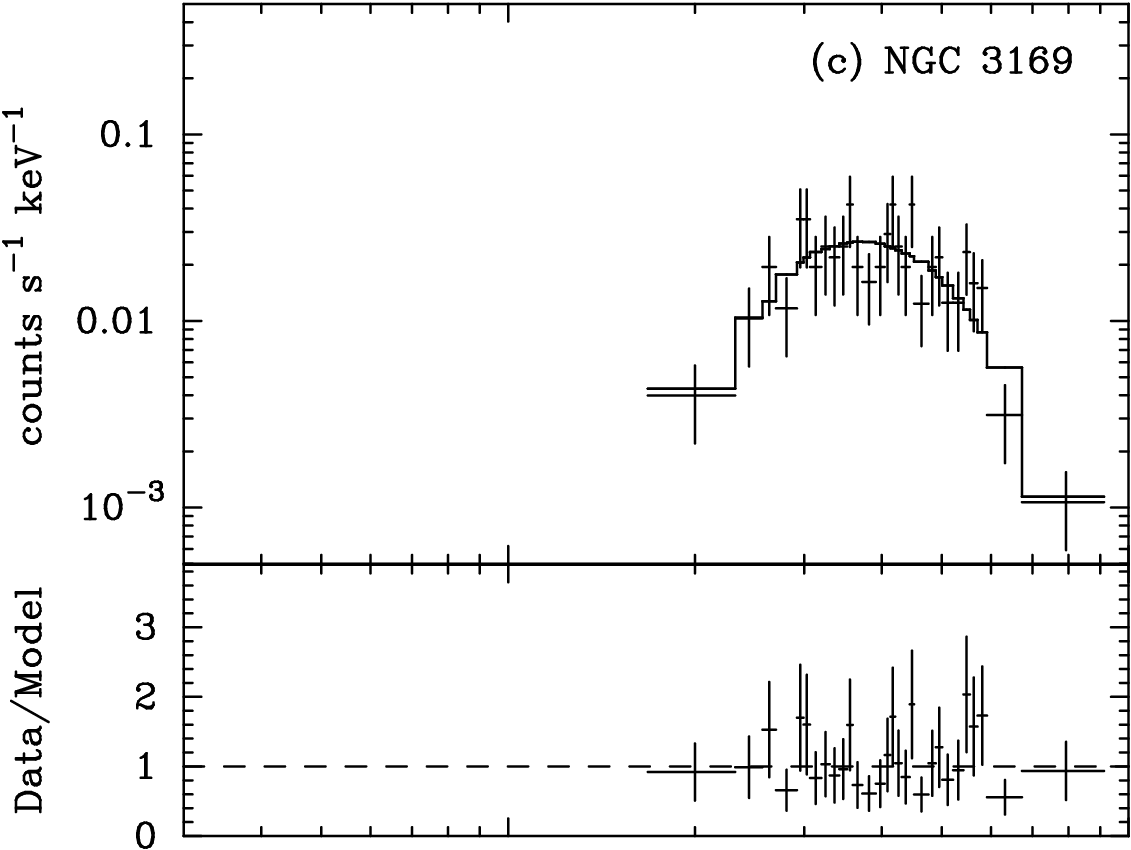}
\includegraphics[scale=0.49,angle=-90]{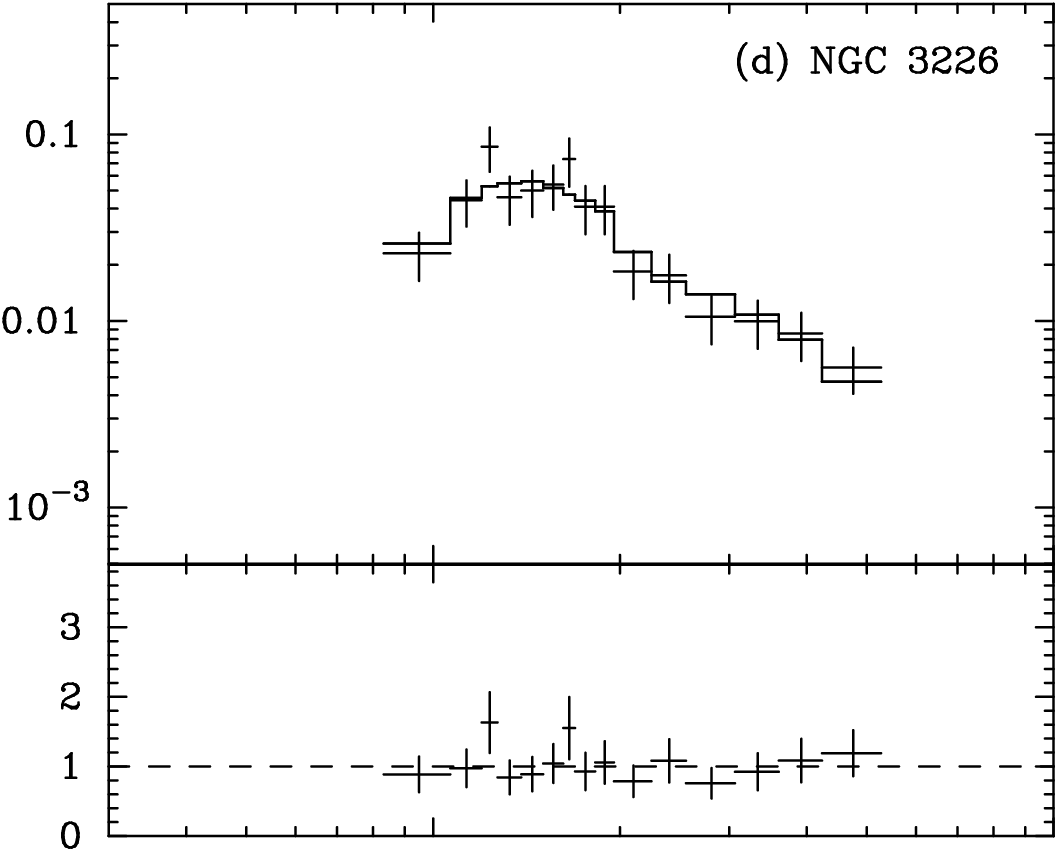}
\includegraphics[scale=0.49,angle=-90]{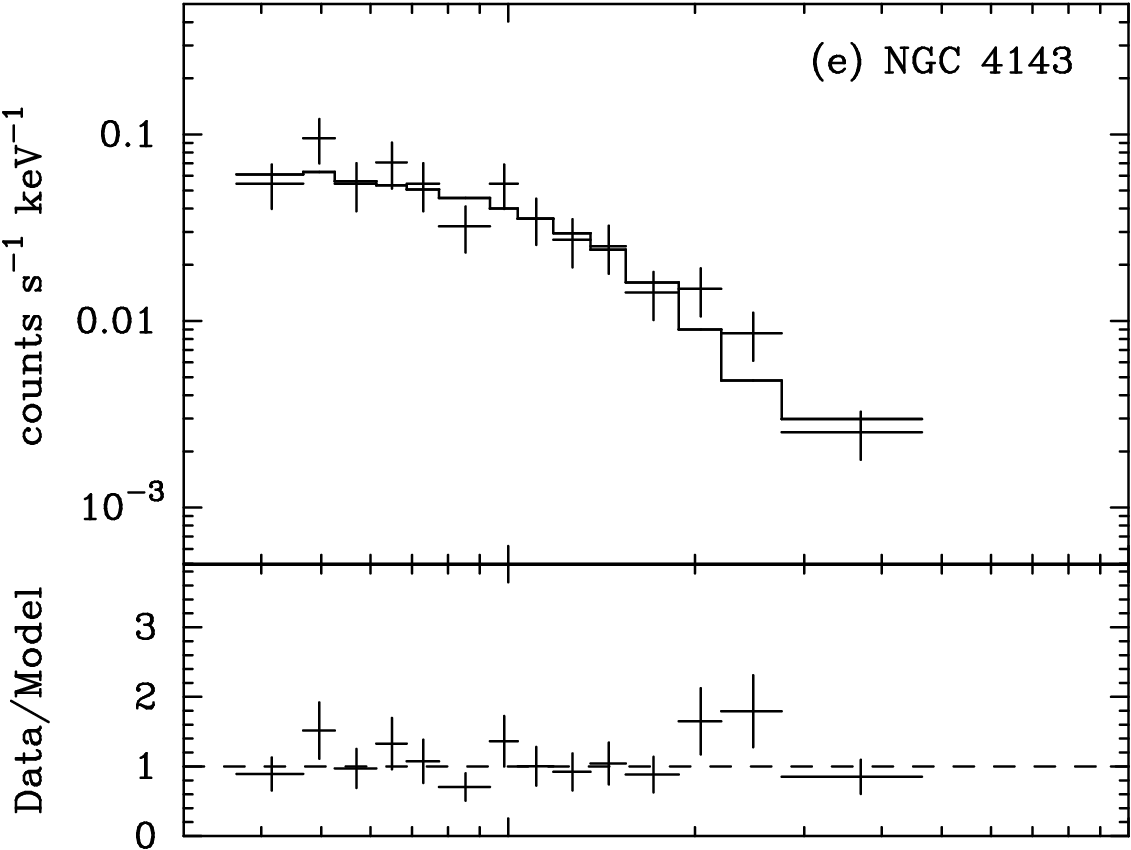}
\hspace{2.2mm} \includegraphics[scale=0.49,angle=-90]{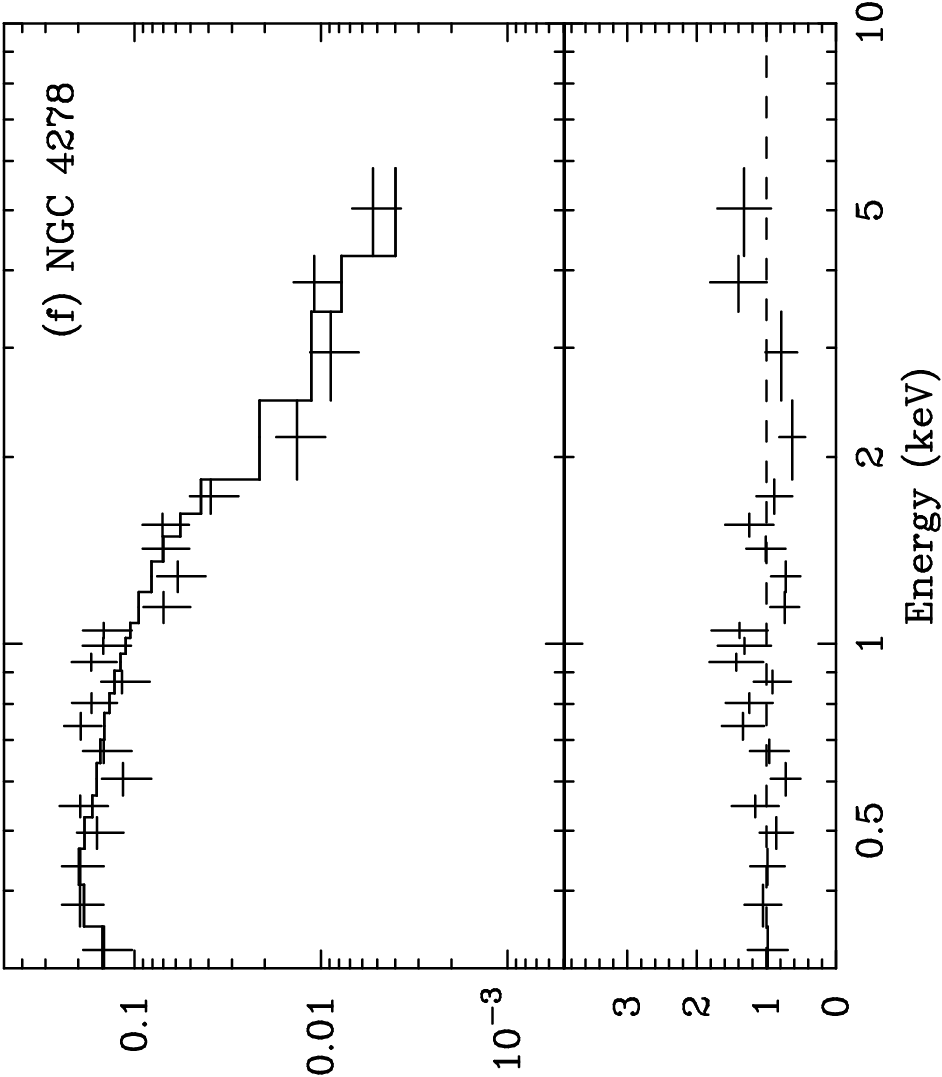}
\figcaption[]{
{\chandra} X-ray Spectra.
({\it a}) NGC 266. 
({\it b}) NGC 3147. 
({\it c}) NGC 3169.
({\it d}) NGC 3226.
({\it e}) NGC 4143.
({\it f}) NGC 4278.
({\it g}) NGC 4548.
({\it h}) NGC 4565.
({\it i}) NGC 6500.
\label{fig-2}
}
\end{figure*}

\setcounter{figure}{1}
\begin{figure*}[t]
\noindent
\includegraphics[scale=0.49,angle=-90]{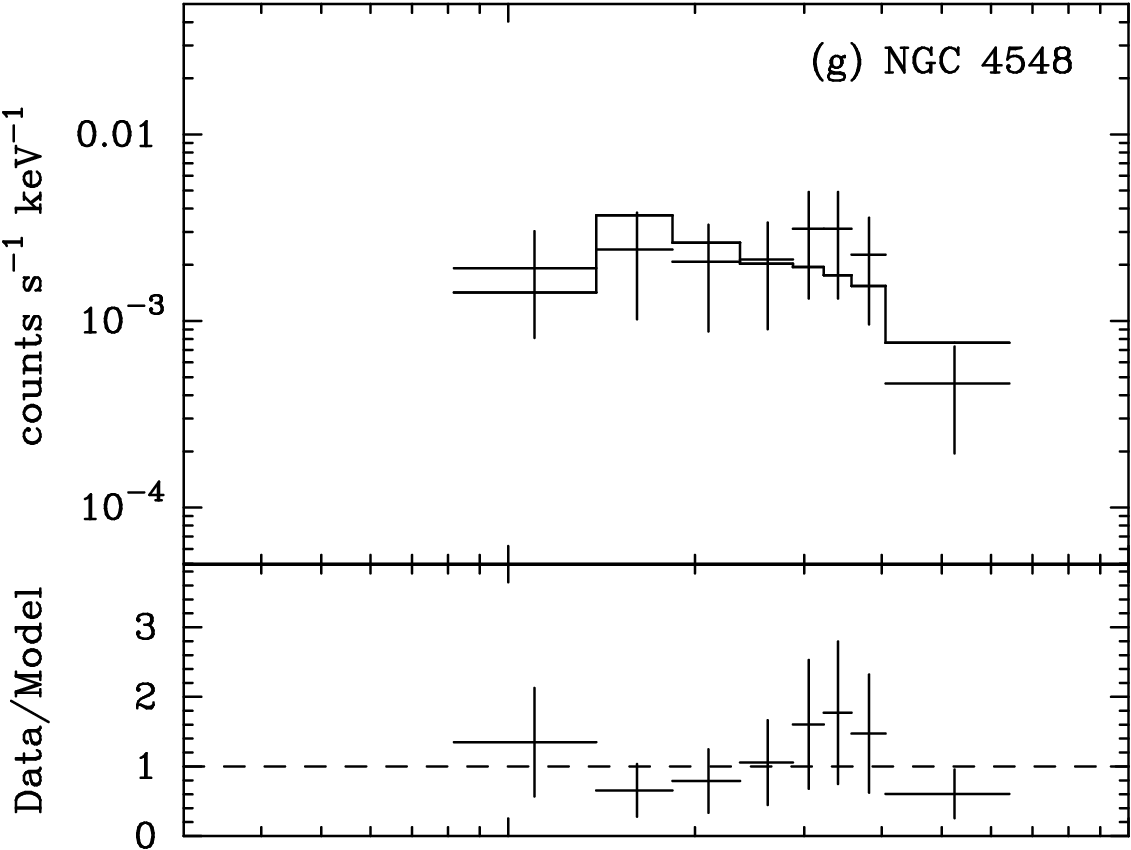}
\includegraphics[scale=0.49,angle=-90]{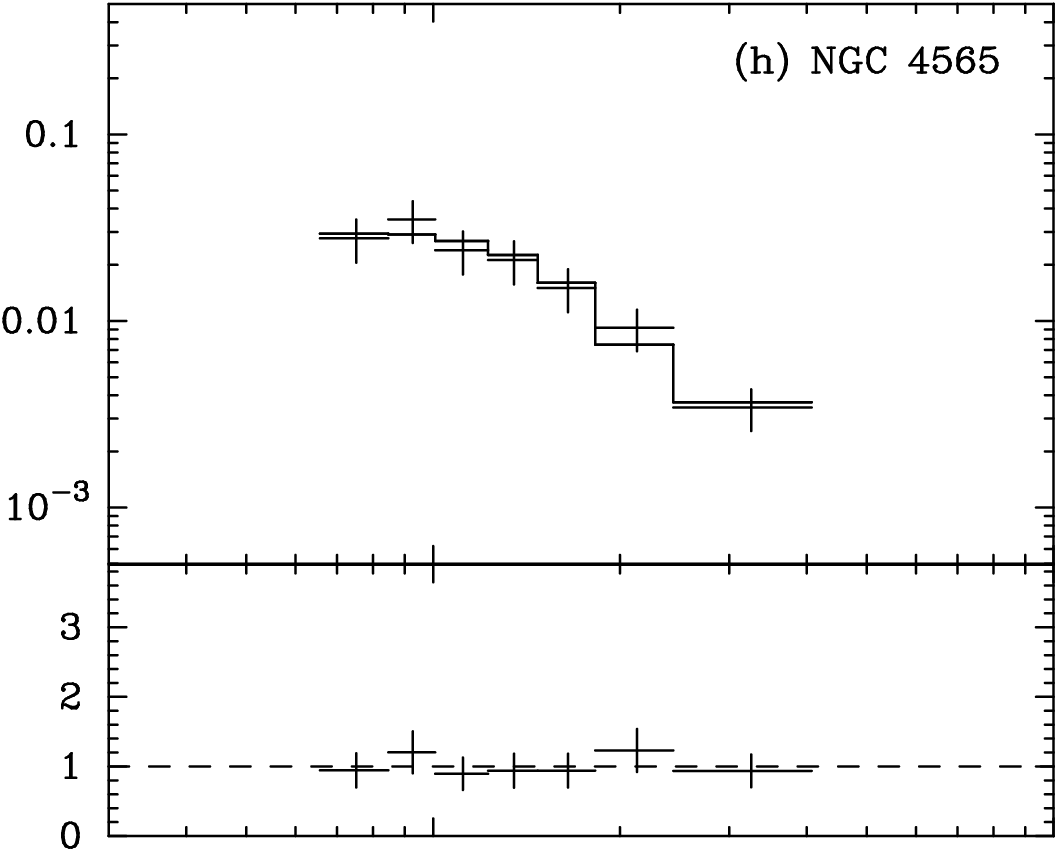}
\includegraphics[scale=0.49,angle=-90]{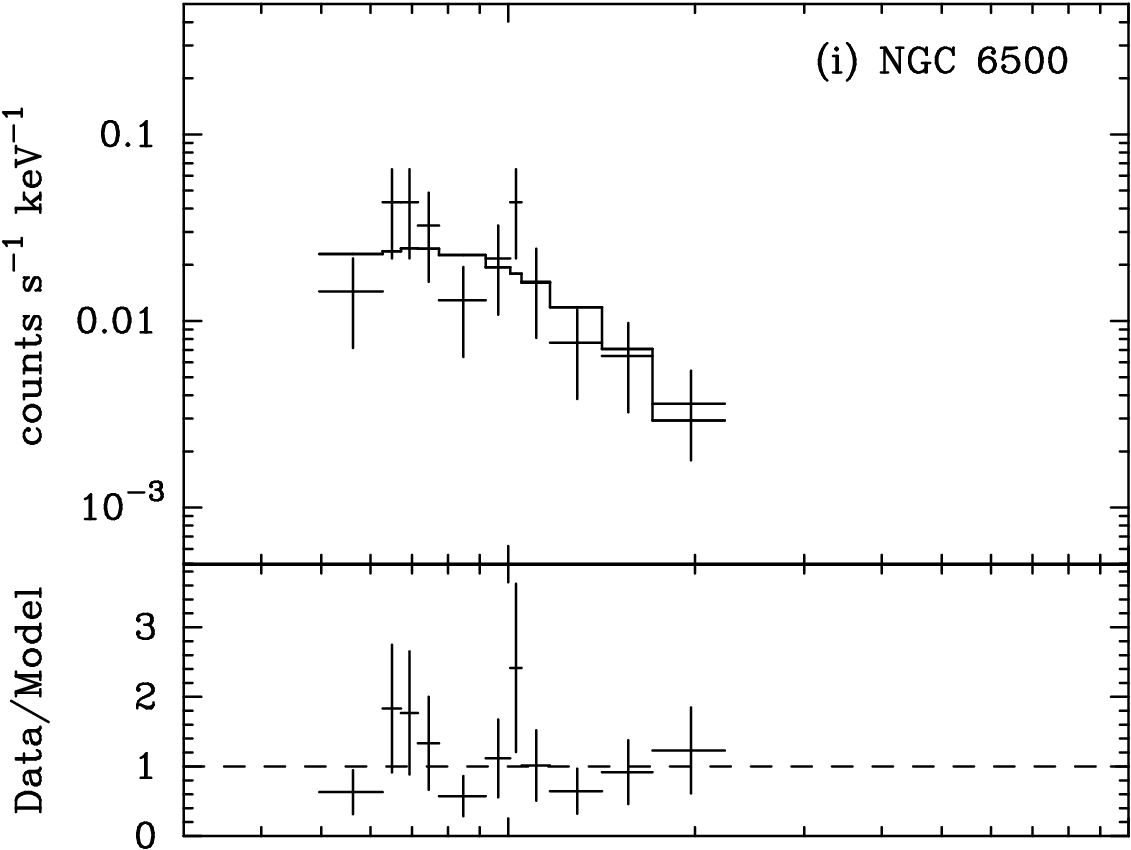}
\figcaption[]{continued.}
\end{figure*}

with fixed parameters to
the spectral models, after normalizing by the ratio of the geometrical
areas of the source and background regions. The errors quoted
represent the 90\% confidence level for one parameter of interest
($\Delta \chi ^2$=2.7).

A power-law model modified by absorption was applied and acceptable
fits were obtained in all cases (Fig. 2). The best-fit parameters for
the nuclear sources are given in Table 4. The observed fluxes and
luminosities (the latter corrected for absorption) in the 2--10 keV
band are shown in Table 2. The results for a few bright off-nuclear
sources are presented in the Appendix. The photon indices of the
nuclear sources are generally consistent with the typical values
observed in LLAGNs (photon index $\Gamma = 1.6-2.0$, e.g., Terashima
et al. 2002a, 2002b), although errors are quite large due to the
limited photon statistics. The spectral slope of NGC 6500
($\Gamma=3.1^{+1.1}_{-1.7}$) is somewhat steeper than is typical of
LLAGNs. This may indicate that there is soft X-ray emission from a
source other than the AGN and/or the intrinsic slope of the AGN is
steep.  One galaxy (NGC 3169, a LINER 2) has a large absorption column
({\NH}=$1.1\times10^{23}$ {\pcm}), while two galaxies (NGC 4548, a
LINER 2, and NGC 3226, a LINER 1.9) show substantial absorption (
$1.6\times10^{22}$ {\pcm}, and $0.9\times10^{22}$ {\pcm},
respectively). Others have small column densities which are consistent
with `type 1' AGNs. No meaningful limit on the equivalent width of an
Fe K$\alpha$ line was obtained for any of the objects because of
limited photon statistics in the hard X-ray band.

One object --- NGC 2787 --- has only 8 detected photons in the 0.5--8
keV band and is too faint to obtain spectral information. A photon
index of 2.0 and the Galactic absorption of $4.3\times10^{20}$ {\pcm}
were assumed to calculate the flux and luminosity which are shown in
Table 2.

\section{Discussion}

\subsection{Power Source of LINERs}

An X-ray nucleus is detected in all the objects except for NGC 4550
and NGC 5866. We test whether the detected X-ray sources are the high
energy extension of the continuum source which powers the optical
emission lines by examining the luminosity ratio {\LX}/{\LHa}. The
H$\alpha$ luminosities ({\LHa}) were taken from Ho et al. (1997a) and
the reddening was estimated from the Balmer decrement for narrow lines
and corrected using the reddening curve of Cardelli, Clayton, \&
Mathis (1989), assuming the intrinsic H$\alpha$/H$\beta$ flux ratio =
3.1. The X-ray luminosities (corrected for absorption)
in the $2-10$ keV band are used.

The H$\alpha$ luminosities and logarithm of the luminosity ratios
{\LX}/{\LHa} are shown in Table 5. The {\LX}/{\LHa} ratios of most
objects are in the range of AGNs ($\log$ {\LX}/{\LHa} $\sim$ 1--2) and
in good agreement with the strong correlation between {\LX} and {\LHa}
for LLAGNs, luminous Seyferts, and quasars presented in Terashima et
al. (2000a) and Ho et al. (2001).  This indicates that their optical
emission lines are predominantly powered by a LLAGN. Note that this
correlation is not an artifact of distance effects, as shown in
Terashima et al. (2000a).

The four objects NGC 2787, NGC 4550, NGC 5866, and NGC 6500, however,
have much lower {\LX}/{\LHa} ratios ($\log$ {\LX}/{\LHa} \simlt 0)
than expected from the correlation ($\log$ {\LX}/{\LHa} $\sim1-2$),
and their X-ray luminosities are insufficient to power the H$\alpha$
emission (Terashima et al. 2000a). This X-ray faintness could indicate
one or more of several possibilities such as (1) an AGN is the power
source, but is heavily absorbed at energies above 2 keV, (2) an AGN is
the power source, but is currently switched-off or in a faint state,
and (3) the optical narrow emission lines are powered by some
source(s) other than an AGN. We briefly discuss these three
possibilities in turn.


If an AGN is present in these X-ray faint objects and absorbed in the
hard energy band above 2 keV, only scattered and/or highly absorbed
X-rays would be observed, and then the intrinsic luminosity would be
much higher than that observed. This can account for the low
{\LX}/{\LHa} ratios and high radio to X-ray luminosity ratios ($\nu
L_{\nu}$(5 GHz)/{\LX}; Table 5 and section 5.3). If the intrinsic
X-ray luminosities are about one or two orders of magnitude higher
than those observed, as is often inferred for Seyfert 2 galaxies
(Turner et al. 1997, Awaki et al. 2000), {\LX}/{\LHa} and $\nu
L_{\nu}$(5 GHz)/{\LX} become typical of LLAGNs.


Alternatively, the AGN might be turned off or in a faint state, with
a higher activity in the past being inferred from the optical emission
lines, whose emitting region is far from the nucleus (e.g., Eracleous
et al. 1995). Also, the radio observations were made a few years
before the {\chandra} ones.  This scenario might thus explain their
relatively low {\LX}/{\LHa} ratios and their relatively high $L_{\rm 5
GHz}$/{\LX} ratios.  If this is the case, the size of the radio core
can be used to constrain the era of the active phase in the recent
past. The upper limits on the size of the core estimated from the beam
size ($\approx$ 2.5 mas) are 0.16, 0.19, and 0.48 pc for NGC 2787, NGC
5866, and NGC 6500, respectively (Falcke et al. 2000). Therefore, the
AGN must have been active until $<$0.52, $<$0.60, and $<$1.6 years,
respectively, before the VLBA observations (made in 1997 June) and
inactive at the epochs (2000 Jan -- 2002 Jan, see Table 1) of the
X-ray observations.  This is an ad hoc proposal and such abrupt
declines of activity are quite unusual, but it cannot be completely
excluded.


It may also be possible that the ionized gas inferred from the optical
emission lines is ionized by some sources other than an AGN, such as
hot stars.  If the observed X-rays reflect the intrinsic luminosities
of the AGN, a problem with the AGN scenario for the three objects NGC
2787, NGC 5866, and NGC 6500 is that these galaxies have very large
$\nu L_{\nu}$(5 GHz)/{\LX} ratios, and would thus be among the radio
loudest LLAGNs. The presence of hot stars in the nuclear region of NGC
6500 is suggested by UV spectroscopy (Maoz et al. 1998). Maoz et
al. (1998) studied the energy budget for NGC 6500 by using the
H$\alpha$ and UV luminosity at 1300 A and showed that the observed UV
luminosity is insufficient to power the H$\alpha$ luminosity even if a
stellar population with the Salpeter initial mass function and a high
mass cutoff of 120$M_{\odot}$ are assumed. This result indicates that
a power source in addition to hot stars must contribute significantly,
and supports the obscured AGN interpretation discussed above.

  The first possibility, i.e., an obscured low-luminosity AGN as the
source of the X-ray emission, seems preferable for NGC 2787, NGC 5866
and NGC 6500, although some other source(s) may contribute to the
optical emission lines.  Additional lines of evidence which support
the presence of an AGN include the fact that all three of these
galaxies (NGC 2787, NGC 5866, and NGC 6500) have VLBI-detected, sub-pc
scale, nuclear radio core sources (Falcke et al. 2000), a broad
H$\alpha$ component (in NGC 2787, and an ambiguous detection in NGC
5866; Ho et al. 1997b), a variable radio core in NGC 2787, and a
jet-like linear structure in a high-resolution radio map of NGC 6500
with the VLBA (Falcke et al. 2000). Only an upper limit to the X-ray
flux is obtained for NGC 5866. If an X-ray nucleus is present in this
galaxy and its luminosity is only slightly below the upper limit, this
source could be an AGN obscured by a column density {\NH}$\sim10^{23}$
{\pcm} or larger. If the apparent X-ray luminosity of the nucleus of
NGC 5866 is {\it much} lower than the observed upper limit, and the
intrinsic X-ray luminosity conforms to the typical {\LX}/{\LHa} ratio
for LLAGN ($\log$ {\LX}/{\LHa} $\approx 1-2$), then the X-ray source
must be almost completely obscured. The optical classification
(transition object) suggests the presence of an ionizing source other
than an AGN, so the low observed {\LX}/{\LHa} ratio could
alternatively be a result of enhanced H$\alpha$ emission
powered by this other ionizing source.

The X-ray results presented above show that the presence of a flat (or
inverted) spectrum compact radio core is a very good indicator of the
presence of an AGN even if its luminosity is very low. On the other
hand, NGC 4550, which does not possess a radio core, shows no evidence
for the presence of an AGN and all the three possibilities discussed
above are viable. If the {\rosat} detection is real (Halderson et
al. 2001), the time variability between the {\rosat} and {\chandra}
fluxes may indicate the presence of an AGN (see Appendix).

It is notable that type 2 LINERs without a flat spectrum compact radio
core may be heterogeneous in nature.  For instance, some LINER 2s
without a compact radio core (e.g., NGC 404 and transition 2 object
NGC 4569) are most probably driven by stellar processes (Maoz et
al. 1998; Terashima et al. 2000b; Eracleous et al. 2002).

\subsection{Obscured LLAGNs}

In our sample, we found at least three highly absorbed LLAGNs (NGC
3169, NGC 3226, and NGC 4548). In addition, if the X-ray faint objects
discussed in section 5.1 are indeed AGNs, they are most probably
highly absorbed with {\NH}$>10^{23}$ {\pcm}. Among these absorbed
objects, NGC 2787 is classified as a LINER 1.9, NGC 3169, NGC 4548,
and NGC 6500 as LINER 2s, and NGC 5866 as a transition 2 object. Thus,
heavily absorbed LINER 2s, of which few are known, are found in the
present observations demonstrating that radio selection is a valuable
technique for finding obscured AGNs. Along with heavily obscured
LLAGNs known in low-luminosity Seyfert 2s (e.g., NGC 2273, NGC 2655,
NGC 3079, NGC 4941, and NGC 5194; Terashima et al. 2002a), our
observations show that at least some type 2 LLAGNs are simply
low-luminosity counterparts of luminous Seyferts in which heavy
absorption is often observed (e.g., Risaliti, Maiolino, \& Salvati
1999). However, some LINER 2s (e.g., NGC 4594, Terashima et al. 2002a;
NGC 4374, Finoguenov \& Jones 2001; NGC 4486, Wilson \& Yang 2002) and
low-luminosity Seyfert 2s (NGC 3147; section 4 and Appendix) show no
strong absorption.  Therefore, the orientation-dependent unified
scheme (e.g., Antonucci 1993) does not always apply to AGNs in the
low-luminosity regime, as suggested by Terashima et al. (2002a).

\subsection{Radio Loudness of LLAGNs}

Combination of X-ray and radio observations is valuable for
investigating a number of areas of AGN physics, including the ``radio
loudness'', the origin of jets, and the structure of accretion
disks. Low-luminosity AGNs (LINERs and low-luminosity Seyfert
galaxies) are thought to be radiating at very low Eddington ratios
($L_{\rm bol}$/$L_{\rm Edd}$) and may possess an advection-dominated
accretion flow (ADAF; see e.g., Quataert 2002 for a recent review). A
study of radio loudness in LLAGNs can constrain the jet production
efficiency by an ADAF-type disk.  Earlier studies have suggested that
LLAGNs tend to be radio loud compared to more luminous Seyferts based
on the spectral energy distributions of seven LLAGNs (Ho 1999) and,
for a larger sample, on the conventional definition of radio loudness
$R_{\rm O}=L_{\nu}$(5 GHz)/$L_{\nu}$(B) (the subscript ``O'', which
stands for optical, is usually omitted but we use it here to
distinguish from $R_{\rm X}$ --- see below), with $R_{\rm O}>10$ being
radio loud (Kellermann et al. 1989, 1994; Visnovsky et al. 1992;
Stocke et al. 1992; Ho \& Peng 2001).  Ho \& Peng (2001) measured the
luminosities of the nuclei by spatial analysis of optical images
obtained with {\HST} to reduce the contribution from stellar light. A
caveat in the use of optical measurements for the definition of radio
loudness is extinction, which will lead to an overestimate of $R_{\rm
O}$ if not properly allowed for.  Although Ho \& Peng (2001) used only
type 1--1.9 objects, some objects of these types show high absorption
columns in their X-ray spectra. In this subsection, we study radio
loudness by comparing radio and hard X-ray luminosities. Since the
unabsorbed luminosity for objects with {\NH} \simgt $10^{23}$ {\pcm}
can be reliably measured in the 2--10 keV band, which is accessible to
{\asca}, {\it XMM-Newton}, and {\chandra}, and such columns correspond
to $A_{\rm V}$ \simgt 50 mag, it is clear that replacement of optical
by hard X-ray luminosity potentially yields considerable
advantages. In addition, the high spatial resolutions of {\it
XMM-Newton} and especially {\chandra} usually allow the nuclear X-ray
emission to be identified unambiguously, while the optical emission of
LLAGN can be confused by surrounding starlight.

In the following analysis, radio data at 5 GHz taken from the
literature are used since fluxes at this frequency are widely
available for various classes of objects.  We used primarily radio
luminosities obtained with the VLA at \simlt $1^{\prime\prime}$
resolution for the present sample. High resolution VLA data at 5 GHz
are not available for several objects. For four such cases, VLBA
observations at 5 GHz with 150 mas resolution are published in the
literature (Falcke et al. 2000) and are used here. For two objects, we
estimated 5 GHz fluxes from 15 GHz data by assuming a spectral slope
of $\alpha=0$ (cf. Nagar et al. 2001). The radio luminosities used in
the following analysis are summarized in Table 5.  Since our sample is
selected based on the presence of a compact radio core, the sample
could be biased to more radio loud objects. Therefore, we constructed
a larger sample by adding objects taken from the literature for which
5 GHz radio, 2--10 keV X-ray, and $R_{\rm O}$ measurements are
available.

First, we introduce the ratio $R_{\rm X} = \nu L_{\nu}$(5 GHz)/{\LX}
as a measure of radio loudness and compare the ratio with the
conventional $R_{\rm O}$ parameter. The X-ray luminosity {\LX} in the
2--10 keV band (source rest frame), corrected for absorption, is used
\footnote{Monochromatic X-ray luminosities can also be used to define
the radio loudness instead of luminosities in the 2--10 keV band; such
a definition would be analogous to $R_{\rm O}$, which utilizes
monochromatic B-band luminosities.  This alternative provides
completely identical results if the X-ray spectral shape is known and
the range of spectral slopes is not large. For example, the conversion
factor $L_{\nu}$(2 keV)/{\LX} is 0.31, 0.26, and 0.22 keV$^{-1}$ for
photon indices of 2, 1.8, and 1.6, respectively, and no
absorption.}. We examine the behavior of $R_{\rm X}$ using samples of
AGN over a wide range of luminosity, including LLAGN, the Seyfert
sample of Ho \& Peng (2001) and PG quasars which are also used in
their analysis. $R_{\rm O}$ parameters and radio luminosities were
taken from Ho \& Peng (2001) for the Seyferts and Kellermann et
al. (1989) for the PG sample.  The values of $R_{\rm O}$ in Kellermann
et al. (1989) have been recalculated by using only the core component of
the radio luminosities.  The optical and radio luminosities of the PG
quasars were calculated assuming $\alpha_{\rm r}=-0.5$ and $\alpha_{\rm
o}=-1.0$ ($S_{\nu}\propto \nu^\alpha$). The X-ray luminosities (mostly
measured with {\asca}) were compiled from Terashima et al. (2002b),
Weaver, Gelbord, \& Yaqoob (2001), George et al. (2000), Reeves \&
Turner (2000), Iwasawa et al. (1997, 2000), Sambruna, Eracleous, \&
Mushotzky (1999), Nandra et al. (1997), Smith \& Done (1996), and
Cappi et al. (1996). Note that only a few objects (NGC 4565, NGC 4579,
and NGC 5033) in our radio selected sample have reliable measurements
of nuclear $L_{\nu}$(B).

Fig. 3. compares the parameters $R_{\rm O}$ and $R_{\rm X}$ for the
Seyferts and PG sample. These two parameters correlate well for most
Seyferts.  Some Seyferts have higher $R_{\rm O}$ values than indicated
by most Seyferts.  This could be a result of extinction in the optical
band. Seyferts showing X-ray spectra absorbed by a column greater than
$10^{22}$ cm$^{-2}$ (NGC 2639, 4151, 4258, 4388, 4395, 5252, and 5674)
are shown as open circles in Fig. 3. At least four of them have a
value of $R_{\rm O}$ larger than indicated by the correlation. The
correlation between $\log R_{\rm O}$ and $\log R_{\rm X}$ for the less
absorbed Seyferts can be described as $\log R_{\rm O}$ = 0.88 $\log
R_{\rm X}$ + 5.0.  According to this relation, the boundary between
radio loud and radio quiet object ($\log R_{\rm O}$ = 1) corresponds
to $\log R_{\rm X} = -4.5$.  The values of $R_{\rm O}$ and $R_{\rm X}$
for a few obscured Seyferts are consistent with the correlation,
indicating that optical extinction is not perfectly correlated with the
absorption column density inferred from X-ray spectra.

The PG quasars show systematically lower $R_{\rm O}$ values than those
of Seyferts at a given $\log R_{\rm X}$.  For the former objects,
$\log R_{\rm O}=1$ corresponds to $\log R_{\rm X}=-3.5$.  This
apparently reflects a luminosity dependence of the shape of the SED:
luminous objects have steeper optical-X-ray slopes $\alpha _{\rm ox} =
1.4-1.7$ ($S\propto\nu^{-\alpha}$; e.g., Elvis et al. 1994, Brandt,
Laor, \& Wills 2000), where $\alpha _{\rm ox}$ is often measured as
the spectral index between 2200 A and 2 keV, while less luminous AGNs have
$\alpha _{\rm ox} = 1.0-1.2$ (Ho 1999).  This is related to the fact
that luminous objects show a more prominent ``big blue bump'' in their
spectra.  Fig. 8 of Ho (1999) demonstrates that low-luminosity objects
are typically 1--1.5 orders of magnitude fainter in the optical band
than luminous quasars for an given X-ray luminosity. Note that none of
the PG quasars used here shows a high absorption column in its X-ray
spectrum below 10 keV.

The definition of radio loudness using the hard X-ray flux ($R_{\rm
X}$) appears to be more robust than that using the optical flux
because X-rays are less affected by both extinction at optical
wavelengths and the detailed shape of the blue bump, as noted
above. Further, measurements of nuclear X-ray fluxes of Seyferts and
LLAGNs with {\chandra} are easier than measurements of nuclear
optical fluxes, since in the latter case the nuclear light must be
separated from the surrounding starlight, a difficult process for
LLAGNs.

Fig. 4 shows the X-ray luminosity dependence of $R_{\rm X}$. In this
plot, the LLAGN sample discussed in the present paper is

\begin{figure*}[t]
\includegraphics[scale=0.55,angle=-90]{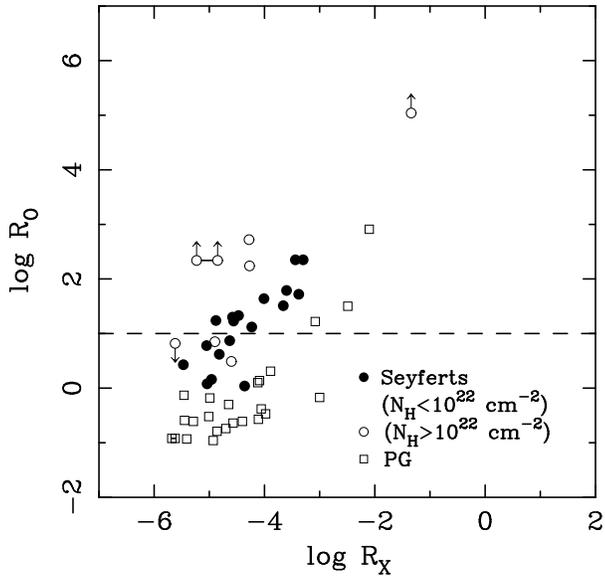}
\figcaption[] { Relation between $R_{\rm O}=L_{\nu}$(5
GHz)/$L_{\nu}$(B) and $R_{\rm X}=\nu L_{\nu}$(5 GHz)/{\LX} for
Seyferts and PG quasars.  The radio luminosity, $L_\nu$(5 GHz),
includes only the nuclear core ($<1^{\prime\prime}$ size) component of
the radio emission. The conventional boundary between ``radio loud''
and ``radio quiet'' objects ($\log R_{\rm O}=1$) is shown as a
horizontal dashed line.  The two open circles connected with a line
correspond to X-ray observations of NGC 4258 at two different epochs.
\label{fig-3}
}
\end{figure*}

\begin{figure*}[t]
\includegraphics[scale=0.55,angle=-90]{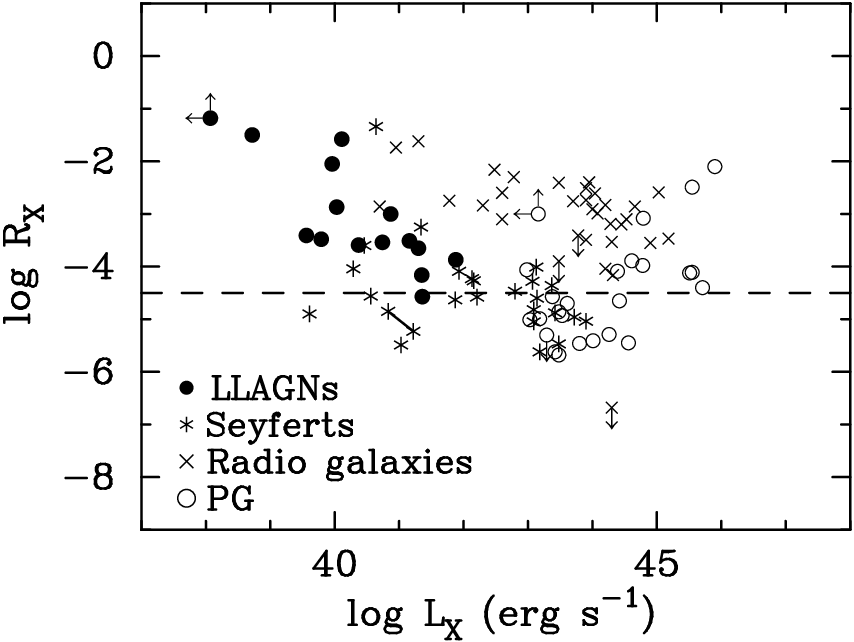}
\includegraphics[scale=0.55,angle=-90]{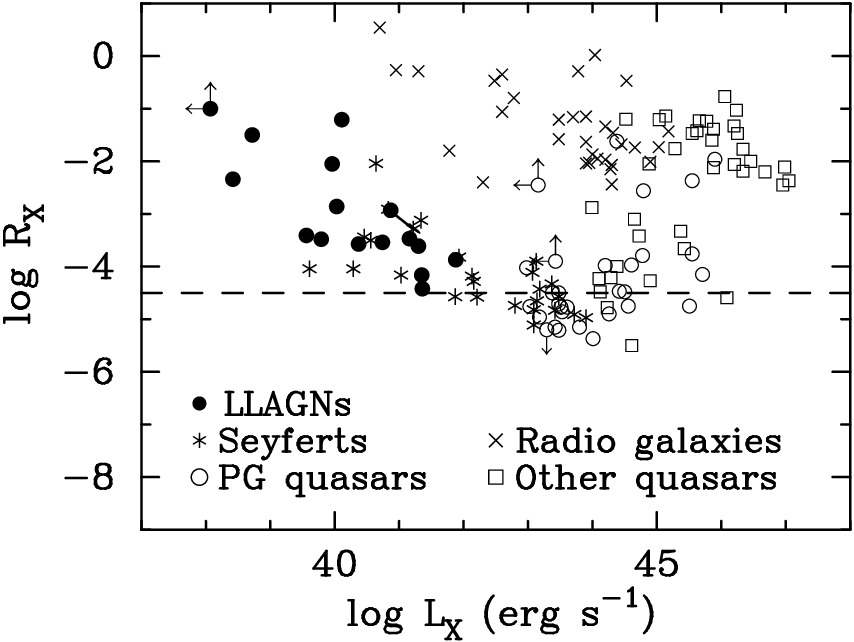}
\figcaption[] { Dependence of $R_{\rm X}=\nu L_{\nu}$(5 GHz)/{\LX} on
{\LX} for the present LLAGN sample, Seyfert galaxies, radio galaxies,
and PG quasars.  The radio luminosity, $L_\nu$(5 GHz), includes only
the nuclear core ($<1^{\prime\prime}$ size) component of the radio emission.
The approximate boundary between ``radio loud'' and ``radio quiet''
objects ($\log R_{\rm x}=-4.5$) is shown as a horizontal dashed
line. The two asterisks connected with a line correspond to X-ray
observations of NGC 4258 at two different epochs.
\label{fig-4}
}
\figcaption[]{ Dependence of $R_{\rm X}=\nu L_{\nu}$(5 GHz)/{\LX} on
{\LX} for the present LLAGN sample, Seyfert galaxies, radio galaxies,
PG quasars, and other quasars, in which {\it total} radio luminosities
(instead of the nucleus-only radio luminosities used in Fig. 4) were
used to calculated $R_{\rm X}$.  The approximate boundary between
``radio loud'' and ``radio quiet'' objects ($\log R_{\rm x}=-4.5$) is
shown as a horizontal dashed line. The two asterisks connected with a
line correspond to X-ray observations of NGC 4258 at two different
epochs.
\label{fig-5}
}
\end{figure*}

\noindent
shown in
addition to the Seyfert and PG samples used above. This is an ``X-ray
version'' of the $\log R_{\rm O}$-$M_{B}^{\rm nuc}$ plot (Fig. 4 in Ho
\& Peng 2001). Radio galaxies taken from Sambruna et al. (1999) are
also added and we use radio luminosities from the core only. Our plot
shows that a large fraction ($\sim70$\%) of LLAGNs ({\LX}$<10^{42}$
{\eps}) are ``radio loud''. This is a confirmation of Ho \& Peng's
(2001) finding. Note, however, that our sample is not complete in any
sense, and this radio-loud fraction should be measured using a more
complete sample. Since radio emission in LLAGNs is likely to be
dominated by emission from jets (Nagar et al. 2001; Ulvestad \& Ho
2001), these results suggest that, in LLAGN, the fraction of the
accretion energy that powers a jet, as opposed to electromagnetic
radiation, is larger than in more luminous Seyfert galaxies and
quasars. Since LLAGNs are thought to have an ADAF-type accretion flow,
such might indicate that an ADAF can produce jets more efficiently
than the geometrically thin disk believed present in more luminous
Seyferts.

The three LLAGNs with the largest $R_{\rm X}$ in Fig. 4 are the three
X-ray faint objects discussed in section 5.1 (NGC 2787, NGC 5866, and
NGC 6500) and which are most probably obscured AGNs.  If their
intrinsic X-ray luminosities are 1--2 orders of magnitude higher than
those observed, their values of $R_{\rm X}$ become smaller by this
factor and are then in the range of other LLAGNs. Even if we exclude
these three LLAGNs, the radio loudness of LLAGNs is distributed over a
wide range: the radio-loudest LLAGNs have $R_{\rm X}$ values similar
to radio galaxies and radio-loud quasars, while some LLAGNs are as
radio quiet as radio-quiet quasars.

A comparison with blazars is of interest to compare our sample with
objects for which the nuclear emission is known to be dominated by a
relativistic jet and thus strongly beamed.  The average $\log R_{\rm
X}$ for high-energy peaked BL Lac objects (HBLs), low-energy peaked BL
Lac objects (LBLs), and flat spectrum radio quasars (FSRQs) are
--3.10, --1.27, and --0.95, respectively, where we used the average
radio and X-ray luminosities for a large sample of blazars given in
Table 3 of Donato et al. (2001). The average $\log R_{\rm X}$ for HBLs
is similar to that for LLAGNs in our sample, while the latter two
classes are about two orders of magnitude more radio loud than
LLAGNs. Although LLAGNs and HBLs have similar values of $\log R_{\rm
X}$, the spectral slope in the X-ray band is different: LLAGNs have a
photon index in the range 1.7--2.0 (see also Terashima et al. 2002),
while HBLs usually show steeper spectra (photon index $>$ 2, e.g.,
Fig. 1 in Donato et al. 2001), and the X-ray emission is believed to
be dominated by synchrotron radiation. Furthermore, blazars with a
lower bolometric luminosity tend to have a synchrotron peak at a
higher frequency and a steeper X-ray spectral slope than higher
bolometric luminosity blazars (Donato et al. 2001).

We also constructed an $R_{\rm X}$-{\LX} plot (Fig. 5) using the {\it
total} radio luminosities of the radio source (i.e. including the
core, jets, lobes, and hot spots, if present). The radio data were
compiled from V\'eron-Cetty \& V\'eron (2001), Kellermann et
al. (1989), and Sambruna et al. (1999). The PG sample and other
quasars are shown with different symbols.  This plot appears similar to
Fig. 4 for LLAGNs, Seyferts, and radio-quiet quasars since these
objects do not possess powerful jets or lobes and off-nuclear radio
emission associated with the AGN is generally of low luminosity
(Ulvestad \& Wilson 1989, Nagar et al. 2001, Ho \& Ulvestad 2001,
Kellermann et al. 1989). On the other hand, radio galaxies have
powerful extended radio emission and consequently the $R_{\rm X}$
values calculated using the total radio luminosities become higher
than if only nuclear luminosities are used. We used the same X-ray
luminosities as in Fig. 4, because jets, lobes, and hot spots are
almost always much weaker than the nucleus in X-rays.  In fact, in our
observations of LLAGNs, we found no extended emission directly related
to the AGN. Thus, the differences between Fig. 4 and Fig. 5 result
from the extended radio emission.

\section{Summary}

Fourteen galaxies with a nuclear radio source having a flat or
inverted spectrum have been observed with {\chandra} with a typical
exposure time of 2 ksec. An X-ray nucleus is detected in all but one
object (NGC 5866). 
11 galaxies have X-ray and H$\alpha$ luminosities
in good accord with the correlation known for AGNs over a wide
range of luminosity, which indicates that these objects are AGNs and
that the AGN is the dominant power source of their optical emission
lines. Their X-ray luminosities are between $5\times10^{38}$ and
$8\times10^{41}$ {\eps}. The three objects NGC 2787, NGC 5866, and NGC
6500 have significantly lower X-ray luminosities than expected
from the {\LX}-{\LHa} correlation. Various observations suggest that
these objects are most likely to be heavily obscured AGNs.  These
observational results show that radio and hard X-ray observations
provide an efficient way to find LLAGN in nearby galaxies, even if the
nuclei are heavily obscured.

One object (the LINER 2 NGC 4550), which does not show a radio core, was
also observed for comparison. No X-ray nucleus is detected. If the
X-ray source detected in this galaxy with {\rosat} is indeed the
nucleus, the nucleus must be variable in X-rays, which would 
indicate the presence of an AGN.

We have used the ratio $R_{\rm X}=\nu L_{\nu}$(5 GHz)/{\LX} as a
measure of radio loudness and found that a large fraction of LLAGNs
are radio loud. This confirms earlier results based on nuclear
luminosities in the optical band, but our results based on hard X-ray
measurements are much less affected by obscuration and the detailed
shape of the ``big blue bump''. We speculate that the increase in
$R_{\rm X}$ as {\LX} decreases below $10^{42}$ {\eps} may result from
the presence of an advection-dominated accretion flow in the inner
part of the accretion flow in low-luminosity objects.  However, the
steep X-ray spectra in our sample of LLAGNs rule out high temperature
thermal bremsstrahlung as the X-ray emission mechanism.

\acknowledgments

Y.T. is supported by the Japan Society for the Promotion of Science
Postdoctoral Fellowship for Young Scientists.  This research was
supported by NASA through grants NAG81027 and NAG81755 to the
University of Maryland.

\appendix


\section{Notes on Individual Objects}

In this Appendix, we compare our results with previously published
results particularly in the hard X-ray band obtained with {\asca} and
{\chandra}. The optical spectroscopic classification is given in
parentheses after the object name.

\noindent {\it NGC 2787 (L1.9)}. --- A result on the same
data set is presented in Ho et al. (2001). Our detected number of
counts is in good agreement with their result.

\noindent {\it NGC 3147 (S2)}. --- This object was observed with
{\asca} in 1993 September and the observed flux was
$1.6\times10^{-12}$ {\eps}{\pcm} in the 2--10 keV band (Ptak et
al. 1996, 1999; Terashima et al. 2002b). Our {\chandra} image is
dominated by the nucleus and shows that the off-nuclear source
contribution within the {\asca} beam is negligible.  Therefore, a
comparison between the observed {\chandra} flux ($3.6\times10^{-12}$
{\eps}{\pcm}), which is 2.3 times larger than that of {\asca}, implies
time variability providing additional evidence for the presence of an
AGN.

In the {\asca} spectrum, a strong Fe-K emission line is detected at
$6.49\pm0.09$ keV (source rest frame) with an equivalent width of
$490^{+220}_{-230}$ eV.  One interpretation of this relatively large
equivalent width is that the nucleus is obscured by a large column
density and the observed X-rays are scattered emission (Ptak et
al. 1996). However, the luminosity ratios {\LX}/{\LHa} and
{\LX}/$L_{\rm [O~III]\lambda 5007}$ suggest small obscuration
(Terashima et al. 2002b).

The observed variability supports the interpretation that the X-ray
emission is not scattered emission from a heavily obscured
nucleus. This galaxy is an example of a Seyfert 2 with only little
absorption in the X-ray band.

\noindent {\it NGC 3226 (L1.9)}. --- This galaxy was observed with
the {\chandra} HETG in 1999 December (George et al. 2001). They obtained
an intrinsic luminosity of $3.2\times10^{40}$
($(2.7-4.6)\times10^{40}$ {\eps}, 68\% confidence limit) in the 2--10
keV band after conversion to a distance of 23.4 Mpc. This luminosity is
consistent with our value of $5.5\times10^{40}$
($(4.6-7.0)\times10^{40}$ {\eps}, 90\% confidence range) after
correction for absorption.

\noindent {\it NGC 4203 (L1.9)}. --- A result on the same data set is
presented in Ho et al. (2001). The nucleus of this object has a large
X-ray flux and pileup is severe in this observation.  A bright source
is seen 2$^{\prime}$ SE of the nucleus which was also separated from
the nucleus with {\asca} SIS observations (Iyomoto et al. 1998;
Terashima et al. 2002b).  The {\chandra} observations show that there
is no source confusing the {\asca} observation of the nucleus.
Therefore, we used an {\asca} flux in the discussions.

\noindent {\it NGC 4278 (L1.9)}. --- A result on the same
data set is presented by Ho et al. (2001). 

We analyzed archival {\asca} data observed on 1998 May 24.  The
effective exposure times after standard data screening were 19.6 ksec
for each SIS and 23.9 ksec for each GIS.  The {\asca} spectrum is well
fitted with a power law with a photon index 1.85 (1.77$-$1.94). The
best-fit absorption column is {\NH}=0, with an upper limit of
$5.4\times10^{20}$ {\pcm}.  The observed flux in the 2$-$10 keV band is
$2.0\times10^{-12}$ {\eps}{\pcm}.

Our {\chandra} image in the hard energy band is dominated by the
nucleus and no bright source is seen in the field. Therefore, the hard
X-ray measurement with {\asca} seems reliable. The {\chandra} flux in
the 2--10 keV band ($7.1\times10^{-13}$ {\eps}{\pcm}) is about
one-third of the {\asca} flux indicating variability.

\noindent {\it NGC 4550 (L2)}. --- This source is not detected with
the present {\chandra} observation. 
A detection with the {\rosat} PSPC is reported by Halderson et
al. (2001).  The observed {\rosat} flux in the 0.1--2.4 keV band is
$1.65\times10^{-13}$ {\eps} {\pcm}. This {\rosat} source is offset
from the optical nucleus by 10$^{\prime\prime}$. If this source is
indeed the nucleus, our non detection by {\chandra} ($F_{\rm
0.1-2.4~keV}<6\times10^{-15}$ {\eps}~{\pcm}) indicates time
variability.

\noindent {\it NGC 4565 (S1.9)}. --- The nuclear region is dominated
by two sources: the nucleus and an off-nuclear source which is
brighter than the nucleus. The observed {\chandra} fluxes of these two
source in the 2--10 keV band ($3.2\times10^{-13}$ and
$5.8\times10^{-13}$ {\eps}{\pcm}) are slightly lower than those
obtained with {\asca} ($5.3\times10^{-13}$ and $1.1\times10^{-12}$
{\eps}{\pcm}; Mizuno et al. 1999; Terashima et al. 2002b),
respectively. These differences appear not to be significant given the
statistical, calibration, and spectral-modeling uncertainties.  (The
uncertainties on the {\chandra} fluxes are dominated by the
statistical errors, which are $\sim30$ \% for the off-nuclear source and
$\sim$50\% for the nucleus, while the error in  the {\asca} fluxes is
dominated by calibration uncertainties of $\sim10$ \%.)

The {\chandra} spectrum of the off-nuclear source can be fitted
by an absorbed power law model with a photon index of
$\Gamma=1.89^{+0.24}_{-0.19}$ and {\NH} = $1.5^{+0.5}_{-0.3}\times10^{21}$
{\pcm}. A multicolor disk blackbody model also provides a good fit
with best-fit parameters $kT_{\rm in}=0.90^{+0.14}_{-0.16}$ keV and
{\NH} = $4.2^{+3.5}_{-4.0}\times10^{20}$ {\pcm}.

\noindent {\it NGC 4579 (L1.9/S1.9)}. --- A result on the same
data set is presented by Ho et al. (2001). The nucleus is significantly
piled up in the {\chandra} observation. The {\chandra} hard band image
is dominated by the nucleus and no bright source is seen in the field.
Therefore, we used {\asca} fluxes observed in 1995 and 1998. Detailed
{\asca} results are published in Terashima et al. (1998, 2000c).

A long (33.9 ksec exposure) {\chandra} observation performed in 2000
May is presented in Eracleous et al. (2002). The 2--10 keV flux
reported is $5.2\times10^{-12}$ {\eps}{\pcm} which is similar to that
of the second {\asca} observation in 1998 ($(5.3-6.1)\times10^{-12}$
{\eps}{\pcm}).

\noindent {\it NGC 5033 (S1.5)}. --- A result on the same data set is
presented in Ho et al. (2001). The nucleus is significantly piled up
in the {\chandra} observation. The five off-nuclear sources shown in
Table 3 are located within the {\asca} beam. The sum of the counts
from these sources is less than 44 counts in the 2--8 keV band, while
380 counts are detected from the nucleus before correction for pileup.
Therefore, the {\asca} flux ($5.5\times10^{-12}$ {\eps}{\pcm};
Terashima et al. 1999, 2002b) is probably larger than the true nuclear
flux by $\approx10$\% or less, unless the off-nuclear sources show
drastic time variability. We used the {\asca} flux without any
correction for the off-nuclear source contribution. The 
10\% uncertainty does not affect any of the conclusions.

We performed a spectral fit to the brightest off-nuclear source
(CXOU~J131329.7+363523). An absorbed power law model was applied, and
{\NH} = 0.40 ($<$1.2) $\times 10^{22}$ {\pcm} and a photon index 
$\Gamma=1.5\pm1.1$ were obtained.

\noindent {\it NGC 5866 (T2)}. ---  
Extended emission of diameter $\approx$ 30$^{\prime\prime}$ ($\sim $2 kpc) 
is seen. The spectrum of this emission may be represented by a
MEKAL plasma model with $kT\approx$ 1 keV and abundance of 0.15 solar.
This component could be identified with a gaseous halo of this S0
galaxy.




\clearpage

\begin{table}[htb]
\begin{center}
	\tablenum{1}
       \caption{Observation log}
\begin{tabular}{cccccccc}
\tableline
\tableline
Name	& $D$ 	& Class		& Date		& Exposure	& \multicolumn{2}{c}{Count rate}	& Notes \\
	& (Mpc)	&		&		& (s)	& (s$^{-1}$)	& (frame$^{-1}$)	& \\
(1)	& (2)	& (3)		& (4)		& (5)		& (6)	& (7)	& (8) \\
\tableline
NGC 266	& 62.4  & L1.9		& 2001 Jun 1	& 2033	& 0.020 &  0.0080	& a \\
NGC 2787& 13.3  & L1.9		& 2000 Jan 7	& 1050	& 0.0075 & 0.024	& c\\
NGC 3147& 40.9  & S2		& 2001 Sep 19	& 2202	& 0.54	& 0.21		& a\\
NGC 3169& 19.7  & L2		& 2001 May 2	& 1953	& 0.081	& 0.033		& a\\
NGC 3226& 23.4  & L1.9		& 2001 Mar 23	& 2228	& 0.094	& 0.038		& a\\
NGC 4143& 17.0  & L1.9		& 2001 Mar 26	& 2514	& 0.063	& 0.025		& a\\
NGC 4203& 9.7   & L1.9		& 1999 Nov 4	& 1754	& 0.17	& 0.54		& c\\
NGC 4278& 9.7   & L1.9		& 2000 Apr 20	& 1396	& 0.18	& 0.33		& b\\
NGC 4548& 16.8  & L2		& 2001 Mar 24	& 2746	& 0.0097& 0.0039	& a\\
NGC 4550& 16.8  & L2		& 2001 Mar 24	& 1885	& ...	& ...		& a\\
NGC 4565& 9.7   & S1.9		& 2000 Jun 30	& 2828	& 0.045	& 0.081		& b\\
NGC 4579& 16.8  & L1.9/S1.9	& 2000 Feb 23	& 2672	& 1.1 & 0.47		& a\\
NGC 5033& 18.7  & S1.5		& 2000 Apr 28	& 2904	& 0.33	& 0.59		& b\\
NGC 5866& 15.3  & T2		& 2002 Jan 10	& 2247	& ...	& ...		& a\\
NGC 6500& 39.7  & L2		& 2000 Aug 1	& 2104	& 0.020	& 0.064		& c\\
\tableline
\end{tabular}
\end{center}

\tablecomments{Col. (1): Galaxy Name. 
Col. (2): Adopted distance taken from Tully (1988).
Col. (3): Spectral class of the nucleus taken from Ho et al. (1997a), where L = LINER, 
S = Seyfert, T = transition objects (LINER/\ion{H}{2}), 1 = type 1, 2 = type 2,
and a fractional number between 1 and 2 denotes various intermediate types.
Col. (4): Observation start date.
Col. (5): Exposure time.
Col. (6): Count rate in the 0.5--8 keV band in units of counts s$^{-1}$ (not corrected 
	for pile up).
Col. (7): Count rate in the 0.5--8 keV band in units of counts per frame readout (not corrected for pile up).
Col. (8): Notes (a)  1/8 chip subarray mode used. (b) 1/2 chip subarray mode used.
(c) full chip mode used.
}
\end{table}


\begin{table}[bht]
\tabletypesize{\small}
\begin{center}
	\tablenum{2}
       \caption{Nuclear Sources}
\begin{tabular}{cccccccccl}
\tableline
\tableline
Name    & RA	& Dec.	& 	& Counts & & Hard/Soft & Flux	& Luminosity	& Notes\\
	& (J2000)& (J2000) & (0.5--8 keV)& (0.5-2 keV) & (2-8 keV) & Band ratio	& (2--10 keV)	& (2--10 keV)\\
(1)	& (2)	& (3)	& (4)	& (5)	& (6)	& (7)	& (8) 	& (9)	& (10) \\
\tableline
NGC 266 & 0 49 47.81 & 32 16 40.0 & 40.7$\pm$6.4 & 29.8$\pm$5.5 & 10.9$\pm$3.3 & 0.37$\pm$0.13 & 1.6	& 7.5	& a\\
NGC 2787&  9 19 18.70 & 69 12 11.3 & 7.9$\pm$2.8 & $<$12.0 & $<$6.4 & ... 			& 0.25  & 0.053	& b\\
NGC 3147& 10 16 53.75 & 73 24 02.8 & 1180.1$\pm$34.4 & 843.4$\pm$29.1 & 333.5$\pm$18.3 & 0.40$\pm$0.03 & 37 & 76& a,c\\
NGC 3169& 10 14 15.05 & 03 27 57.9 & 159.0$\pm$12.6 & $<$11.8 & 151.0$\pm$12.3 & $>$12.9 	& 24	& 26 	& a\\
	&		&		&	&		&	&	& 26	& 22	& a,d\\
NGC 3226& 10 23 27.01 & 19 53 55.0 & 209.3$\pm$14.5 & 125.2$\pm$11.2 & 80.5$\pm$9.0 & 0.64$\pm$0.09 & 7.6 & 5.5	& a\\
NGC 4143& 12  9 36.07 & 42 32 03.0 & 157.4$\pm$12.6 & 121.3$\pm$11.1 & 32.6$\pm$5.7 & 0.27$\pm$0.05 & 3.1 & 1.1 & a\\
NGC 4203& 12 15 05.02 & 33 11 49.9 & 294.3$\pm$17.2 & 198.7$\pm$14.1 & 91.7$\pm$9.6 & 0.46$\pm$0.06 & ...& ...	& e\\
NGC 4278& 12 20 06.80 & 29 16 51.6 & 255.6$\pm$16.4 & 209.6$\pm$14.9 & 52.0$\pm$7.4 & 0.25$\pm$0.04 & 8.1& 0.91	& a,c\\
NGC 4548& 12 35 26.46 & 14 29 46.7 & 26.6$\pm$5.2 & 8.7$\pm$3.0 & 17.7$\pm$4.2 & 2.02$\pm$0.85 & 1.6	& 0.61	& a\\
NGC 4550& ...		& ...	& $<4.6$		& $<4.7$	& $<4.7$	& ...	& $<0.077$ & $<0.026$ 	& \\
NGC 4565& 12 36 20.78 & 25 59 15.7 & 127.3$\pm$11.3 & 92.5$\pm$9.6 & 34.9$\pm$5.9 & 0.38$\pm$0.08 & 3.2	& 0.36	& a\\ 
NGC 4579& 12 37 43.52 & 11 49 05.4 & 3067.9$\pm$55.6 & 2240.5$\pm$47.7 & 812.3$\pm$28.5 & 0.36$\pm$0.01 & ...& ... & e\\
NGC 5033& 13 13 27.47 & 36 35 38.1 & 946.5$\pm$30.9 & 562.1$\pm$23.8 & 380.2$\pm$19.5 & 0.68$\pm$0.05 & ... &...	& e\\
NGC 5866& ...		& ...	& $<4.6$		& $<4.8$	& $<3.0$	& ...	& $<0.064$	& $<0.018$	& b\\
NGC 6500& 17 55 59.78 & 18 20 18.0 & 42.4$\pm$6.6 & 41.5$\pm$6.5 & $<$7.6 & $<$0.18 & 0.28	& 0.55	& a\\
	&		&		&	&		&	&	& 0.69	& 1.3	& a,d\\
\tableline
\end{tabular}
\end{center}
\tablecomments{Col. (1): Galaxy Name. 
Cols. (2--3): X-ray position (units of right ascension are hours, minutes, and seconds, and units of declination are degrees, arcminutes, and arcseconds).
Cols. (4--6): Source counts in 0.5--8 keV, 0.5--2 keV, and 2--8 keV, respectively, not corrected for pile up. Errors are 1$\sigma$.
Col. (7): Hard/Soft band ratio. Errors are 1$\sigma$.
Col. (8): Flux in the 2--10 keV band in units of $10^{-13}$ ergs s$^{-1}$ cm$^{-2}$ not corrected for absorption.
Col. (9): Luminosity in the 2--10 keV band in units of $10^{40}$ ergs s$^{-1}$ corrected for absorption.
Col. (10): Notes. (a) Flux and luminosity measured by spectral fits presented in Table 4.
(b) Photon index = 2 and the Galactic absorption are assumed to calculate the flux and luminosity.
(c) Flux and luminosity corrected for slight pileup effect.
(d) Photon index = 2 is assumed to calculate the flux and luminosity.
(e) Significantly piled up.
}
\end{table}

\clearpage


\begin{deluxetable}{ccccccccccl}
\tablewidth{17.5cm}
\tabletypesize{\scriptsize}
	\tablenum{3}
        \tablecaption{Detected Off-Nuclear Sources}
\tablehead{
\colhead{Name}    & 
\colhead{RA}	& 
\colhead{Dec.}	& 
\colhead{CXOU Name}	&  
\multicolumn{3}{c}{Counts}  & 
\colhead{Hard/Soft} & \colhead{Flux}	& 
\colhead{Luminosity}	&
\colhead{Notes}\\
\colhead{}	& 
\colhead{(J2000)}& 
\colhead{(J2000)}&
\colhead{}	& 
\colhead{(0.5--8 keV)}& 
\colhead{(0.5-2 keV)} & 
\colhead{(2-8 keV)} & 
\colhead{Band ratio}	& 
\colhead{(2--10 keV)}	& 
\colhead{(2--10 keV)}\\
(1)	& (2)	& (3)	& (4)	& (5)	& (6)	& (7)	& (8) 	& (9)	& (10) & \colhead{(11)}\\
}
\startdata
NGC 2787&  9 19 23.05 & 69 14 24.4 & J091923.1+691424 & 21.0$\pm$4.6 & 17.9$\pm$4.2 & $<$8.0 & $<$0.44 & 0.68 & ...	& a,b\\
NGC 3147& 10 16 51.50 & 73 24 08.9 & J101651.5+732409 & 6.7$\pm$2.6 & $<$6.8 & $<$9.2 & ... 		& 0.10	& 0.20	& b\\
NGC 3169& 10 14 14.35 & 03 28 10.8 & J101414.3+032811 & 6.9$\pm$2.6 & 6.9$\pm$2.6 & $<$3.0 & $<$0.43 & 0.11	& 0.051	& b\\
	& 10 14 17.90 & 03 28 55.2 & J101417.9+032855 & 10.9$\pm$3.3 & 8.0$\pm$2.8 & $<$9.4 & $<$1.18 & 0.18	& 0.084	& b\\
NGC 3226& 10 23 26.69 & 19 54 06.8 & J102326.7+195407 & 8.9$\pm$3.0 & 7.9$\pm$2.8 & $<$4.7 & $<$0.59 & 0.13	& 0.085	& b\\
NGC 4203& 12 15 09.20 & 33 09 54.7 & J121509.2+330955 & 240.1$\pm$15.5 & 196.4$\pm$14.0 & 40.8$\pm$6.4 & 0.21$\pm$0.04 &...&...	& a,c, TON 1480\\
	& 12 15 14.33 & 33 11 04.7 & J121514.3+331105 & 11.9$\pm$3.5 & $<$6.3 & 9.9$\pm$4.3 & $>$1.57 & 2.0	& ...	& a,d J121514.3+331105, g\\
	& 12 15 15.34 & 33 13 54.0 & J121515.3+331354 & 6.0$\pm$2.4 & 6.0$\pm$2.4 & $<$3.0 & $<$0.50 	& 0.10	& ...	& a,b\\
	& 12 15 15.64 & 33 10 12.3 & J121515.6+331012 & 16.9$\pm$5.2 & 14.9$\pm$5.0 & $<$6.3 & $<$0.42 	& 0.30	& ...	& a,b, star\\
	& 12 15 19.84 & 33 10 12.2 & J121519.8+331012 & 15.9$\pm$4.0 & 10.9$\pm$3.3 & $<$4.7 & $<$0.43 	& 0.29	& ...	& a,b\\
NGC 4550& 12 35 21.30 & 12 14 04.5 & J123521.3+121405 & 6.0$\pm$2.4 & 5.9$\pm$2.4 & $<$3.0 & $<$0.50 & 0.10	& ...	& a,b\\
	& 12 35 27.76 & 12 13 38.9 & J123527.8+121339 & 35.8$\pm$6.0 & 26.6$\pm$5.2 & 8.0$\pm$2.8 & 0.30$ \pm$0.12 & 1.1&11000	& a,d, QSO 1232+125, h\\
NGC 4565& 12 36 14.65 & 26 00 52.5 & J123614.7+260052 & 14.9$\pm$5.0 & 8.9$\pm$4.1 & 6.0$\pm$3.6 & 0.67$\pm$0.51 	& 0.76	& 0.086	& d, A30, i\\
	& 12 36 17.40 & 25 58 55.5 & J123617.4+255856 & 269.5$\pm$16.4 & 209.7$\pm$14.5 & 59.9$\pm$7.7 & 0.29$\pm$0.04 & 5.8 & 0.67& e, f, A32, i\\
	& 12 36 18.64 & 25 59 34.6 & J123618.6+255935 & 8.8$\pm$3.0 & $<$13.5 & $<$6.2 & ... & 0.098		& 0.011	& b\\
	& 12 36 19.02 & 25 59 31.5 & J123619.0+255932 & 6.9$\pm$2.6 & $<$10.7 & $<$6.2 & ... & 0.077		& 0.009	& b\\
	& 12 36 19.03 & 26 00 27.0 & J123619.0+260027 & 19.9$\pm$4.5 & 18.0$\pm$4.2 & $<$6.4 & $<$0.36 & 0.22	& 0.025	& b, A33, i\\
	& 12 36 20.92 & 25 59 26.7 & J123620.9+255927 & 5.9$\pm$2.4 & $<$10.6 & $<$4.7 & ... 		& 0.065	& 0.007	& b\\
	& 12 36 27.39 & 25 57 32.7 & J123627.4+255733 & 15.9$\pm$4.0 & 14.9$\pm$3.9 & $<$4.8 & $<$0.32 & 0.18	& 0.020	& b,A37?, i\\
	& 12 36 28.12 & 26 00 00.9 & J123628.1+260001 & 12.9$\pm$3.6 & 11.0$\pm$3.3 & $<$6.4 & $<$0.58 & 0.14	& ...	& a,b\\
	& 12 36 31.28 & 25 59 36.9 & J123631.3+255937 & 12.0$\pm$4.6 & $<$18.9 & $<$4.8 & ... & 0.13		& ...	& a,b, A43?, i\\
NGC 5033& 13 13 24.78 & 36 35 03.7 & J131324.8+363504 & 13.9$\pm$3.7 & 10.0$\pm$3.2 & $<$9.4 & $<$0.94 & 0.15	& 0.063	& b\\
	& 13 13 28.88 & 36 35 41.0 & J131328.9+363541 & 6.7$\pm$2.6 & $<$10.2 & $<$6.3 & ... 	& 0.072		& 0.030	& b\\
	& 13 13 29.46 & 36 35 17.3 & J131329.5+363517 & 34.6$\pm$5.9 & 31.7$\pm$5.7 & $<$7.9 & $<$0.25 & 0.37	& 0.16	& b\\
	& 13 13 29.66 & 36 35 23.1 & J131329.7+363523 & 47.7$\pm$6.9 & 31.8$\pm$5.7 & 15.9$\pm$5.1 & 0.50$\pm$0.18 & 2.0& 0.89& f\\
	& 13 13 35.56 & 36 34 04.4 & J131335.6+363404 & 7.0$\pm$2.6 & 6.0$\pm$2.4 & $<$4.8 & $<$0.80 & 0.075	& 0.031	& b\\
NGC 6500& 17 56 01.59 & 18 20 22.6 & J175601.6+182023 & 7.0$\pm$2.6 & $<$10.4 & $<$6.2 & ... 		& 0.12	& 0.23	& b\\
\enddata
\tablecomments{Col. (1): Galaxy Name. 
Cols. (2--3): X-ray position (units of right ascension are hours, minutes, and seconds, and units of declination are degrees, arcminutes, and arcseconds).
Col. (4): CXO source name.
Cols. (5--7): Source counts in 0.5--8 keV, 0.5--2 keV, and 2--8 keV, respectively, not corrected 
for pile up. Errors are 1$\sigma$.
Col. (8): Hard/Soft band ratio. Errors are 1$\sigma$.
Col. (9): Flux in the 2--10 keV band in units of $10^{-13}$ ergs s$^{-1}$ cm$^{-2}$ not corrected for absorption.
Col. (10): Luminosity in the 2--10 keV band in units $10^{40}$ ergs s$^{-1}$ corrected for absorption, where the source is assumed to be within the host galaxy.
Col. (11): Notes and possible identifications. 
(a) Spatially outside the host galaxy.
(b) Flux and luminosity measured by assuming a photon index of 2 and the Galactic absorption.
(c) Significantly piled up.
(d) Flux and luminosity measured by assuming the Galactic absorption and 
power law spectra, with photon indices determined from the band ratio.
(e) Flux and luminosity corrected for slight pileup effect.
(f) Flux and luminosity measured by spectral fits.
(g) Radio counterpart given in Ho \& Ulvestad (2001).
(h) Luminosity calculated by assuming $z=0.723$, $H_{0}$=75 km~s$^{-1}$~Mpc$^{-1}$, and $q_0$=0.5.
(i) {\it ROSAT} source presented in Vogler, Pietsch, \& Kahabka (1996)
}
\end{deluxetable}



\clearpage

\begin{table}[htb]
\begin{center}
	\tablenum{4}
       \caption{Spectral fits}
\begin{tabular}{ccccc}
\tableline
\tableline
Name	& $N_{\rm H}$	& $\Gamma$	& C-statistic (dof)	& Notes\\ 
	& ($\times10^{22}$ cm$^{-2}$)\\
\tableline

NGC 266	& $0.15 (<0.82)$	& $1.40^{+1.8}_{-0.98}$ & 10.6 (6)\\
NGC 3147& $0.148^{+0.037}_{-0.018}$	& $1.79^{+0.17}_{-0.09}$ & 45.4 (41)	& a\\
NGC 3169& $11.2^{+4.6}_{-3.8}$	& $2.6^{+1.2}_{-1.0}$	& 21.3 (26)\\
NGC 3226& $0.93^{+0.40}_{-0.36}$& $2.21^{+0.59}_{-0.55}$ & 8.1 (12)\\
NGC 4143& $0.015 (<0.096)$	& $1.66^{+0.47}_{-0.27}$ & 12.2 (11)\\
NGC 4278& $0 (<0.035)$	& $1.64^{+0.28}_{-0.14}$ & 16.9 (18)& a\\
NGC 4548& $1.6^{+2.6}_{-1.5}$	& $1.7^{+1.9}_{-1.6}$	& 3.7 (5)\\
NGC 4565& $0.095 (<0.38)$	& $1.52^{+0.85}_{-0.57}$ & 1.6 (4)\\
NGC 6500& $0.21 (<0.69)$	& $3.1^{+1.1}_{-1.7}$	& 8.5 (8) \\
\tableline
\end{tabular}
\end{center}
\tablecomments{(a): Corrected for pileup effect.
}
\end{table}


\begin{table}[htb]
\begin{center}
	\tablenum{5}
       \caption{Luminosity Ratios}
\begin{tabular}{cccccc}
\tableline
\tableline
Name	& $\log${\LHa} & $\log${\LX}/{\LHa}	& $\log \nu L_{\nu}$(5 GHz)	& $\log \nu L_{\nu}$(5 GHz)/{\LX} & Notes\\
	& (erg s$^{-1}$) & 			& (erg s$^{-1}$)	& 				& \\
	& (1)	& (2)	& (3)	& (4) 	& (5)\\
\tableline
NGC 266  & 39.36 & 1.52  &	37.87 & $-$3.00 & a\\
NGC 2787 & 38.56 & 0.16 &	37.22 & $-$1.50 & b\\
NGC 3147 & 40.02 & 1.86 &	38.01 & $-$3.87 & c\\
NGC 3169 & 39.52 & 1.82 &	37.19 & $-$4.16 & a\\
NGC 3226 & 38.93 & 1.81 &	37.20 & $-$3.54 & a\\
NGC 4143 & 38.69 & 1.34 &	37.16 & $-$2.87 & b\\
NGC 4203 & 38.35 & 2.02 &	36.79 & $-$3.59 & b, e\\
NGC 4278 & 39.20 & 0.76 &	37.91 & $-$2.05 & b\\
NGC 4548 & 38.48 & 1.31 &	36.31 & $-$3.48 & d\\
NGC 4550 & 38.50 & $<-$0.09& 	36.07 & $>-$2.34 & d\\
NGC 4565 & 38.46 & 1.10 & 	36.15 & $-$3.41 & b\\
NGC 4579 & 39.48 & 1.82 & 	37.65 & $-$3.59 & b, e\\
NGC 5033 & 39.70 & 1.67 & 	36.79 & $-$4.57 & c, e\\
NGC 5866 & 38.82 & $<-$0.56&	36.89 & $>-$1.18 & b\\
NGC 6500 & 40.48 & $-$0.37& 	38.90 & $-$1.21 & a\\
\tableline
\end{tabular}
\end{center}
\tablecomments{Col. (1): Logarithm of narrow H$\alpha$ luminosity corrected for reddening.
Col. (2): Logarithm of the ratio of intrinsic X-ray (2--10 keV band)
to reddening corrected narrow H$\alpha$ luminosity.
Col. (3): Logarithm of radio luminosity ($\nu L_{\nu}$) at 5 GHz.
Col. (4): Logarithm of the ratio of radio to X-ray luminosity.
Col. (5): Notes (a) VLBA data (2 mas resolution) taken from Falcke et al. (2000). (b) VLA data (0.5$^{\prime \prime}$ resolution) taken from Nagar et al. (2001). (c) VLA data (1$^{\prime \prime}$ resolution) taken from Ho \& Ulvestad (2001). (d) VLA data (0.15$^{\prime \prime}$ resolution) at 2 cm taken from Nagar et al. (2000) with the assumption $\alpha=0$. (e) {\it ASCA} X-ray fluxes taken from Terashima et al. (2002a) are used.
}
\end{table}

\end{document}